\documentclass[a4paper,11pt]{article}
\pdfoutput=1
\usepackage{jheppub} 
\usepackage{graphicx,color,rotating}
\usepackage{hyperref}
\usepackage{epsfig,color}
\usepackage{slashed}
\usepackage{amsfonts}
\usepackage{ulem}
\newcommand{\lsim}{\raisebox{-0.13cm}{~\shortstack{$<$ \\[-0.07cm] $\sim$}}~}
\newcommand{\gsim}{\raisebox{-0.13cm}{~\shortstack{$>$ \\[-0.07cm] $\sim$}}~}

\usepackage{tabularx}
\usepackage{graphicx}
\usepackage{adjustbox}
\usepackage{tabularx}

\begin{document}
\title{Correlated gravitational wave and microlensing signals of macroscopic dark matter}

\renewcommand{\thefootnote}{\arabic{footnote}}

\author{
Danny Marfatia$^{1}$ and
Po-Yan Tseng$^{2}$}
\affiliation{
$^1$ Department of Physics \& Astronomy, University of Hawaii at Manoa,
Honolulu, HI 96822, USA \\
$^2$ Department of Physics and IPAP, Yonsei University,
Seoul 03722, Republic of Korea \\
}
\date{\today}

\abstract{
Fermion dark matter particles can aggregate to form extended dark matter structures via a first-order phase transition in 
which the particles get trapped in the false vacuum. We study {\it Fermi balls} created in a phase transition
induced by a generic quartic thermal effective potential. We show that for Fermi balls of mass, $3\times10^{-12}M_\odot \lsim M_{\rm FB} \lsim 10^{-5}M_\odot$, correlated observations of gravitational waves produced during the phase transition (at SKA/THEIA/$\mu$Ares), 
and gravitational microlensing caused by Fermi balls (at Subaru-HSC), can be made.}

\maketitle

\section{Introduction}

The identity of dark matter (DM) is a long-standing puzzle in particle physics, astrophysics and cosmology.
Weakly interacting massive particles are popular DM candidates that attain
the measured DM relic density through thermal freeze-out, and typically
have masses of $\mathcal{O}(10-10^3)~{\rm GeV}$
and weak scale annihilation cross sections.
However, no convincing evidence of these particles has been found over several decades of experimentation. 

Recently, a paradigm-altering connection between DM and first-order phase transitions (FOPTs)
in early universe has garnered attention. In the Standard Model (SM), 
both the electroweak and quantum chromodynamics phase transitions are smooth crossovers,
so the FOPT must occur in a dark sector. In this work, 
we consider a scenario in which a quartic thermal effective potential  gives rise to a FOPT,
and a Yukawa interaction with fermion DM  generates a nonzero DM mass in the true vacuum, 
whereas the DM particle remains massless in the false vacuum.
%
If the DM mass in the true vacuum 
is larger than the critical temperature of the FOPT, then, four-momentum conservation
 causes the DM to be trapped in the false vacuum. If a DM-antiDM asymmetry exists, then as the false vacuum shrinks, 
the DM particles are compressed to form macroscopic objects called {\it Fermi balls} (FBs), which
 become the DM relic~\cite{Hong:2020est}. Similar ideas have been proposed  in Refs.~\cite{Witten:1984rs,Bai:2018dxf}.

In this paper, we study FBs produced in a FOPT generated by a general quartic thermal potential. Our focus is the mass range of FBs for which 
gravitational wave and microlensing signals can be correlated.
%
The Subaru Hyper Suprime-Cam (HSC) sky survey has observed  about 100 million stars in the M31 galaxy
in an observation time of 7 hours, and plans for a 70 hour observation period are underway~\cite{Niikura:2017zjd}.
As a FB passes between a star in M31 and the Earth, the transient brightening of the star by gravitational
micolensing can be detected by Subaru-HSC. Also, future telescopes like SKA~\cite{SKA}, THEIA~\cite{THEIA} and $\mu$Ares~\cite{muAres} will have the ability
to detect gravitational waves from the FOPT that produced the FBs.
%

This paper is organized as follows. We investigate the formation and properties of FBs
in section~\ref{sec:FB_formation}.
In section~\ref{sec:microlensing}, we compute
the microlensing event rate for several benchmark points, and the sensitivity of the Subaru-HSC survey
for the case of extended sources and lenses. In section~\ref{gws}, we calculate the gravitational wave spectra expected from the
FOPT for our benchmark points.
Finally, we summarize in section~\ref{sec:summary}.

\bigskip

\section{Fermi ball formation}
\label{sec:FB_formation}

We consider a scenario in which the dark sector only couples to the SM sector gravitationally. The model is composed of a dark Dirac fermion $\chi$, a dark scalar $\phi$, and their Yukawa interaction:
\begin{eqnarray}
\label{eq:lagrangian}
\mathcal{L}\supset \bar{\chi} i \slashed{\partial} \chi -g_\chi \phi \bar{\chi} \chi -V_{\rm eff}(\phi,T) \,,
\end{eqnarray}
where the last term is the finite-temperature effective potential
of $\phi$ that induces the FOPT in the early universe.
When the temperature drops below the critical temperature $T_c$,
the universe starts to traverse from the false vacuum ($\langle \phi \rangle=0$) 
to the true vacuum ($\langle \phi \rangle = v_\phi$).
The interaction term $\phi \bar{\chi} \chi$ in the Lagrangian 
implies the $\chi$ is massless 
in the false vacuum, and obtains mass $m_\chi\simeq g_\chi v_\phi$ 
in the true vacuum. 
For $\chi$ to acquire mass in the true vacuum, energy conservation dictates that
$\chi$ in the false vacuum have enough kinetic energy 
to penetrate the bubble wall during the FOPT.
Conversely, if
\begin{eqnarray}
m_\chi \simeq g_\phi v_\phi \gg T_c\,,
\end{eqnarray} 
the $\chi$'s will be trapped inside the false vacuum.
As the true vacuum expands and the false vacuum shrinks,
 the $\chi$'s  aggregate and form a macroscopic FB.
 For this to occur, there must be a nonzero asymmetry $\eta_\chi \equiv (n_\chi-n_{\bar{\chi}})/s$ in the number
 densities in the false vacuum (where $s$ is the entropy density) during the phase transition so that an excess remains after pair annihilation $\bar{\chi}\chi \to \phi \phi$, and 
$\chi$ must carry a conserved global $U(1)_Q$ so that the FB attains stability by accumulating 
$Q$-charge~\cite{Hong:2020est}. Mechanisms that produce $\eta_\chi$ are discussed in the appendix of Ref.~\cite{Hong:2020est}.

\bigskip

\subsection{Effective potential}

\begin{figure}[t!]
\centering
\includegraphics[height=2.9in,angle=270]{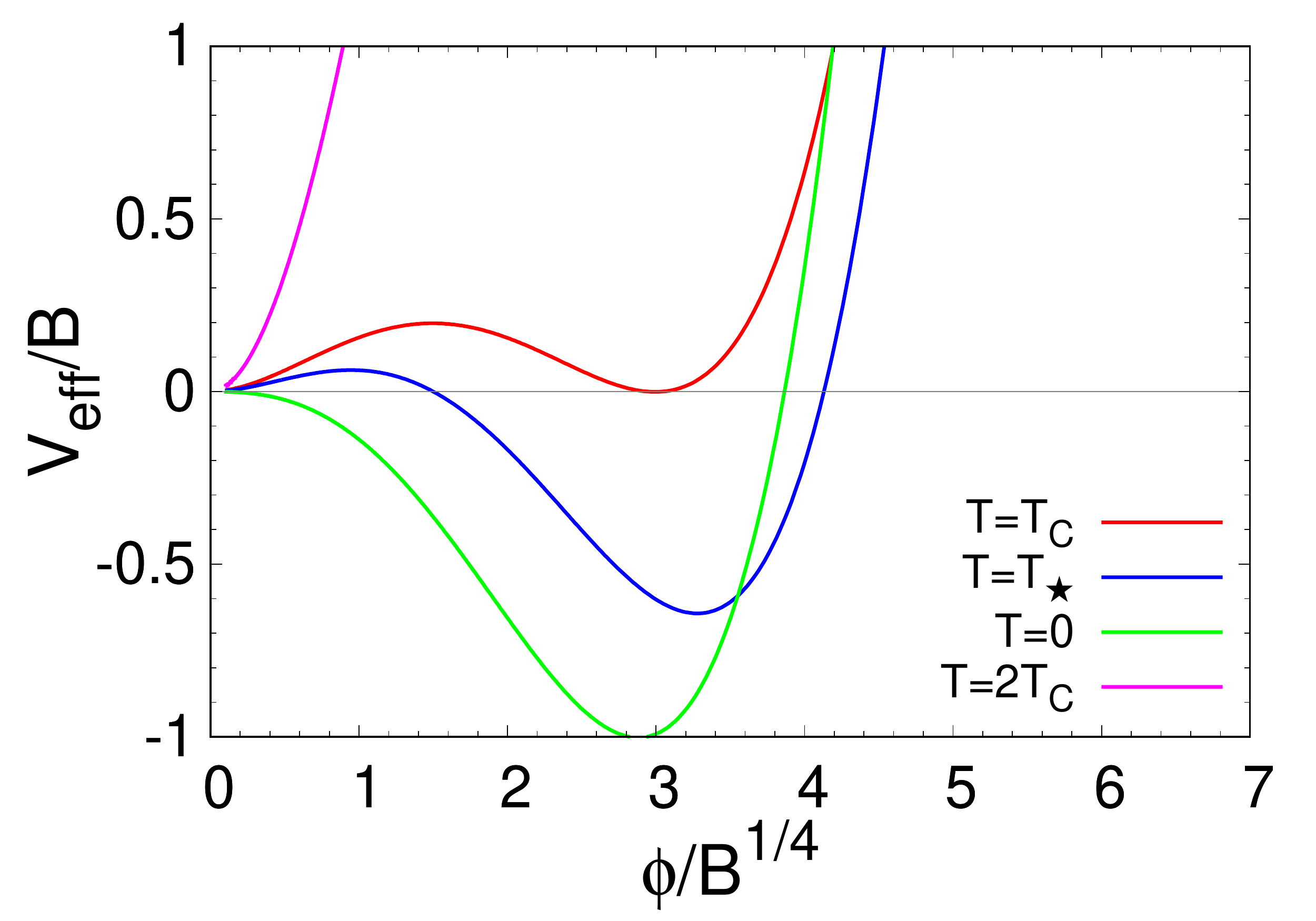}
\includegraphics[height=2.9in,angle=270]{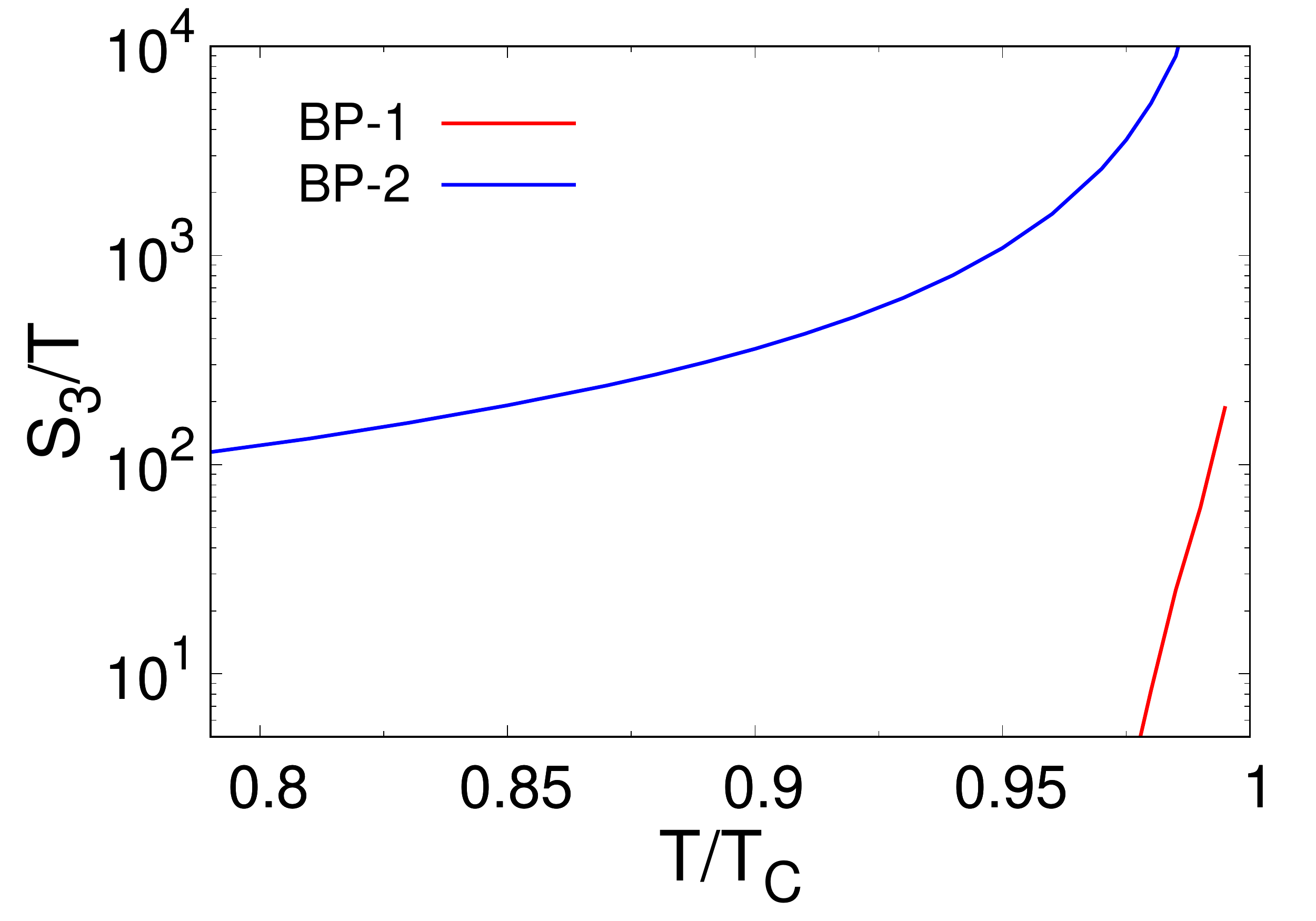}
\caption{\small \label{fig:Veff}
Left-panel: the effective potential $V_{\rm eff}(\phi,T)$ for {\bf BP-2}: $\lambda=0.16$, $A=0.1$, $B=(43.5~{\rm keV})^4$, $C=6.23~{\rm keV}$, $D=0.45$. In this case,
 $T_c= 0.93 B^{1/4}$ and $T_\star=0.79 B^{1/4}$.
Right-panel: $S_3/T$ for benchmark points
{\bf BP-1} and {\bf BP-2} in Table~\ref{2HDM}.
}
\end{figure}



We consider the finite-temperature 
quartic effective potential~\cite{Dine:1992wr,Adams:1993zs},
\begin{eqnarray}
V_{\rm eff}(\phi,T)= D(T^2-T^2_0)\phi^2-(AT+C)\phi^3+\frac{\lambda}{4}\phi^4\,,
\label{eq:V_tree}
\end{eqnarray}
where $T_0$ is the destabilization temperature, $C$ contributes a zero-temperature cubic term,
and $D$, $A$, and $\lambda$ are dimensionless parameters.
Potentials of this form are commonly found in particle physics including inert singlet, inert doublet, minimal supersymmetry, and Majoron models.
At zero temperature, the potential has its global minimum at 
$\tilde{\phi}_\pm=(3C \pm \sqrt{9C^2+8\lambda DT^2_0})/(2\lambda)$ with vacuum energy density, 
\begin{equation}
V_{\rm eff}(\tilde{\phi}_+,0)=-\left(\frac{DT^2_0}{2}+\frac{C}{4}\tilde{\phi}_+ \right)\,{\tilde{\phi}_+}^2\equiv -B\,.
\end{equation}
Therefore, $B$ is the difference in vacuum energy density between the $\phi=0$ and $\tilde{\phi}_+$ phases, 
and will make an important contribution to the latent heat released during the FOPT.
%
%
In terms of the input parameters, 
$$\lambda,~A,~B,~C,~D\,,$$
$T_0$ is a derived quantity.
We show an example of the finite-temperature potential 
in the left panel of Fig.~\ref{fig:Veff}, where the critical temperature is defined by $V_{\rm eff}(0,T_c)=V_{\rm eff}(v_\phi(T_c),T_c)$.

The Euclidean action $S_3(T)/T$ that determines the bubble nucleation rate per unit volume  is given by
\begin{eqnarray}
S_3(T)=4\pi \int^{\infty}_{0} r^2 dr \left[ \frac{1}{2}\left( \frac{d\phi}{dr} \right)^2 +V_{\rm eff}(\phi,T) \right]\,,
\end{eqnarray}
where $\phi$ satisfies the equation of motion,
\begin{eqnarray}
\frac{d^2\phi}{dr^2}+\frac{2}{r}\frac{d\phi}{dr}=\frac{\partial V_{\rm eff}(\phi,T)}{d\phi}\,,
\end{eqnarray}
with boundary conditions,
\begin{eqnarray}
\frac{d\phi}{dr}\vert_{r=0}=0 \ \ \ ~\text{and \ \ \ $\phi(r\to \infty)=0$}\,. 
\end{eqnarray}
%

An analytical approximation for $S_3$ is available for quartic potentials of the form,
\begin{equation}
V_{\rm eff}(\phi,T) \simeq \bar{\lambda} \phi^4-a\phi^3+b\phi^2\,,
\end{equation}
where the coefficients are temperature dependent. 
Reference~\cite{Adams:1993zs} finds
\begin{eqnarray}
S_3(T)=\frac{\pi a}{\bar{\lambda}^{3/2}} \frac{8\sqrt{2}}{81}
(2-\delta)^{-2} \sqrt{\delta/2}
(\beta_1 \delta + \beta_2 \delta^2 + \beta_3 \delta^3)\,,
\end{eqnarray}
where $\delta\equiv 8\bar{\lambda} b/a^2$, $\beta_1=8.2938$, $\beta_2=-5.5330$, and $\beta_3=0.8180$.
For illustration, $S_3/T$ for benchmark points
{\bf BP-1} and {\bf BP-2}  in Table~\ref{tab:BP} are shown 
in the right-panel of Fig.~\ref{fig:Veff}.

In terms of the bubble nucleation rate per unit volume,
\begin{eqnarray}
\label{eq:nucleation_rate}
\Gamma(T)=T^4 \left( \frac{S_3}{2\pi T} \right)^{3/2} e^{-\frac{S_3}{T}}\,,
\end{eqnarray}
the fraction of space in the false vacuum ($\langle \phi \rangle=0$) is
\begin{eqnarray}
\label{eq:F_t}
F(t)={\rm exp}\left[ -\frac{4\pi }{3} v^3_w\int^{t}_{t_c} dt' (t-t')^3 \Gamma(t') \right]\,,
\end{eqnarray}
where $v_w$ is the bubble wall velocity and $t_c$ is the time corresponding to the critical temperature. Note that $t$ and $T$ are related by the 
Hubble parameter $H$:
\begin{eqnarray}
dt = -{\frac{dT}{T H(T)}}\,,~~~~~
H^2(T) =  \frac{\rho(T)}{3 M^2_{\rm Pl}}\,,
\end{eqnarray}
where the radiation energy density, $\rho(T) = \frac{\pi^2}{30} g_\ast T_{\rm SM}^4$, with $g_\ast = g^{\rm SM}_\ast +g^{\rm D}_\ast\left(T/T_{\rm SM}\right)^4$  the total number of relativistic degrees of freedom when the dark sector is at temperature $T$ and corresponding 
temperature of the SM sector is $T_{\rm SM}$. Note that $g^{\rm D}_\ast =4.5$ at all relevant times.

The latent heat converted to dark radiation during the phase transition at $T=T_\star$ is
\begin{equation}
\epsilon(T_\star) = \left(1-T\frac{\partial}{\partial T}\right)\Delta V|_{T_\star}\,,
\end{equation}
where $\Delta V = V_{\rm eff}(0,T)-V_{\rm eff}(v_\phi(T),T)$. This injection of energy changes the temperature of the dark sector from $T_\star$ to
$T_{f}$ (after the phase transition), which is given by
\begin{equation}
 \frac{\pi^2 g^{\rm D}_\ast }{30}  T_{f}^4 =  \frac{\pi^2 g^{\rm D}_\ast}{30}  T_{\star}^4+ \epsilon(T_\star)\,.
\end{equation}

The effective number of extra neutrino species contributed by the dark sector after the phase transition, $\Delta N_{\rm eff}$, depends sensitively on $T_{f}/T_{{\rm SM}\star}$. 
For example, for temperatures below 60~keV, $g^{\rm SM}_\ast \simeq 3.36$, and $\Delta N_{\rm eff} \simeq 9.9 (T_{f}/T_{{\rm SM}\star})^4$. The 95\%~C.L. upper bound from a combination of cosmic microwave background, baryon acoustic oscillations and Big Bang Nucleosynthesis (BBN) measurements that uses the primordial helium abundance of Ref.~\cite{Izotov:2014fga} is $\Delta N_{\rm eff} < 0.55$~\cite{Planck:2018vyg}.
Also, the Hubble tension suggests 
that at recombination $\Delta N_{\rm eff}$ is in the range $0.4-1$~\cite{Riess:2016jrr}. A robust 95\% C.L. upper bound from BBN alone is $\Delta N_{\rm eff} < 1$~\cite{Mangano:2011ar}. Absent knowledge of the reheating process after inflation,  we select benchmark points with values of $T_{f}/T_{{\rm SM}\star}$ that give $\Delta N_{\rm eff} \leq 0.5$.

%

We identify $T_\star$ with the temperature of percolation, i.e., 
the temperature at which the fraction of space remaining in the false vacuum is $1/e$.
Then $T_\star$ and the corresponding time $t_\star$ are given 
by the condition,
\begin{eqnarray}
F(t_\star)=1/e\simeq 0.37 \,.
\label{cond}
\end{eqnarray}
We also take $t_\star$ to be the time when FBs form.  

For phase transitions much shorter than the Hubble time, the inverse duration of the phase transition is
\begin{eqnarray}
\beta\equiv {\dot{\Gamma} \over \Gamma} \simeq \left. -\frac{d(S_3/T)}{dt} \right|_{t=t_\star}\,,
\label{beta}
\end{eqnarray}
which is often expressed as
\begin{equation}
{\beta \over H_{\star}} \simeq T_{\star} \left.{d(S_3/T) \over dT}\right|_{T_{\star}} \,  .
\end{equation}
In addition to this parameter, the GW spectrum depends on the strength of the phase transition,
\begin{equation}
\alpha \equiv  \frac{\epsilon(T_\star)}{\rho(T_\star)}\,.
\end{equation}
Because the fraction of latent heat converted to GWs is determined by the dynamics in the dark sector (and is not related to $\alpha$)~\cite{Nakai:2020oit}, we can assume it is unity.

\bigskip

\subsection{Number density and density profile of Fermi balls}

FBs start to form at $T_\star$ in the false vacuum,
as it shrinks and separates into smaller volumes.
Below a critical volume  of the {\it false vacuum bubble}, $V_\star=4\pi R^3_\star/3$, bubble nucleation of the true vacuum stops and the
formation of FBs takes over.
Since the timescale on which the false vacuum bubble shrinks is $\Delta t = R_\star/v_w$, 
$V_\star$ is given by
$\Gamma(T_\star)V_\star \Delta t\sim 1$.
Then, with one FB per critical volume, 
the number density of FBs $n_{\rm FB}|_{T_\star}$ is determined by
$n_{\rm FB}|_{T_\star} V_\star = F(t_\star)$, i.e.,~\cite{Hong:2020est}:
\begin{eqnarray}
\label{eq:nFB}
n_{\rm FB}|_{T_\star}= \left( \frac{3}{4\pi} \right)^{1/4}
\left( \frac{\Gamma(T_\star)}{v_w} \right)^{3/4} F(t_\star)\,.
\end{eqnarray}

%
%
The net $Q$-charge trapped in a FB is given by~\cite{Hong:2020est}
\begin{eqnarray}
Q_{\rm FB}= \eta_\chi 
\left( \frac{ s}{n_{\rm FB}}\right)_{T_\star}\,,
\end{eqnarray}
which is equivalent to the total number of $\chi$ that form a FB.
We assume all the $\chi$'s are trapped in the false vacuum
and cannot penetrate the bubble wall due to its large mass in the true vacuum.
The $\chi$-asymmetry $\eta_\chi$ is 
a free parameter that must be tuned to produce the measured DM relic density.
Because the universe evolves adiabatically, $n_{\rm FB}/s$ and $Q_{\rm FB}$ remain unchanged,
and today,
\begin{eqnarray}
\label{eq:nFB_today}
n_{\rm FB}|_0=s_0 \left(\frac{n_{\rm FB}}{s}\right)_{T_\star} \,,
\end{eqnarray}
where the temperature of the universe today is $T_{{\rm SM}0}=0.235~{\rm meV}$ ($\simeq 0$ for our purposes), and the total entropy density of the universe today is
\begin{equation}
s_0 = {2\pi^2 \over 45} \bigg[ g_{*s}^{\rm SM}(T_{{\rm SM}0})+ g_{*s}^{\rm D}  {g_{*s}^{\rm SM}(T_{{\rm SM}0})\over g_{*s}^{\rm SM}(T_{{\rm SM\star}})}  \big({T_f \over T_{\rm SM\star}}\big)^3\bigg] T_{{\rm SM}0}^3\,.
\end{equation}
Here, $g_{*s}$ is the number of relativistic degrees of freedom for the entropy density, and $g_{*s}^{\rm D}=4.5$ at all relevant times. The dark sector contribution to the entropy density is significantly suppressed compared to that of the SM sector.


To find the density profile of FBs
we need to solve the  Tolman-Oppenheimer-Volkoff (TOV) equation~\cite{Douchin:2001sv}.
The energy of a FB is~\cite{Hong:2020est}
\begin{eqnarray}
\label{eq:FB_E}
E= \frac{3\pi}{4} \left( \frac{3}{2\pi} \right)^{2/3} 
\frac{Q^{4/3}_{\rm FB}}{R}
+ 4\pi \sigma_0 R^2 + \frac{4\pi}{3}B R^3\,,
\end{eqnarray}
where the first term is the Fermi-gas pressure of $\chi$ in a FB, 
$\sigma_0$ is the surface tension, and $B =  V_{\rm eff}(0, 0)-V_{\rm eff}(v_\phi,  0)$.
Thus the energy density becomes
\begin{eqnarray}
\rho_{\rm FB}(n_{\chi})=\left( \frac{9\pi}{8} \right)^{2/3} n^{4/3}_{\chi}+B\,,
\end{eqnarray}
with $n_{\chi}$ the number density of $\chi$ in a FB.

\begin{figure}[t!]
\centering
\includegraphics[height=2.9in,angle=270]{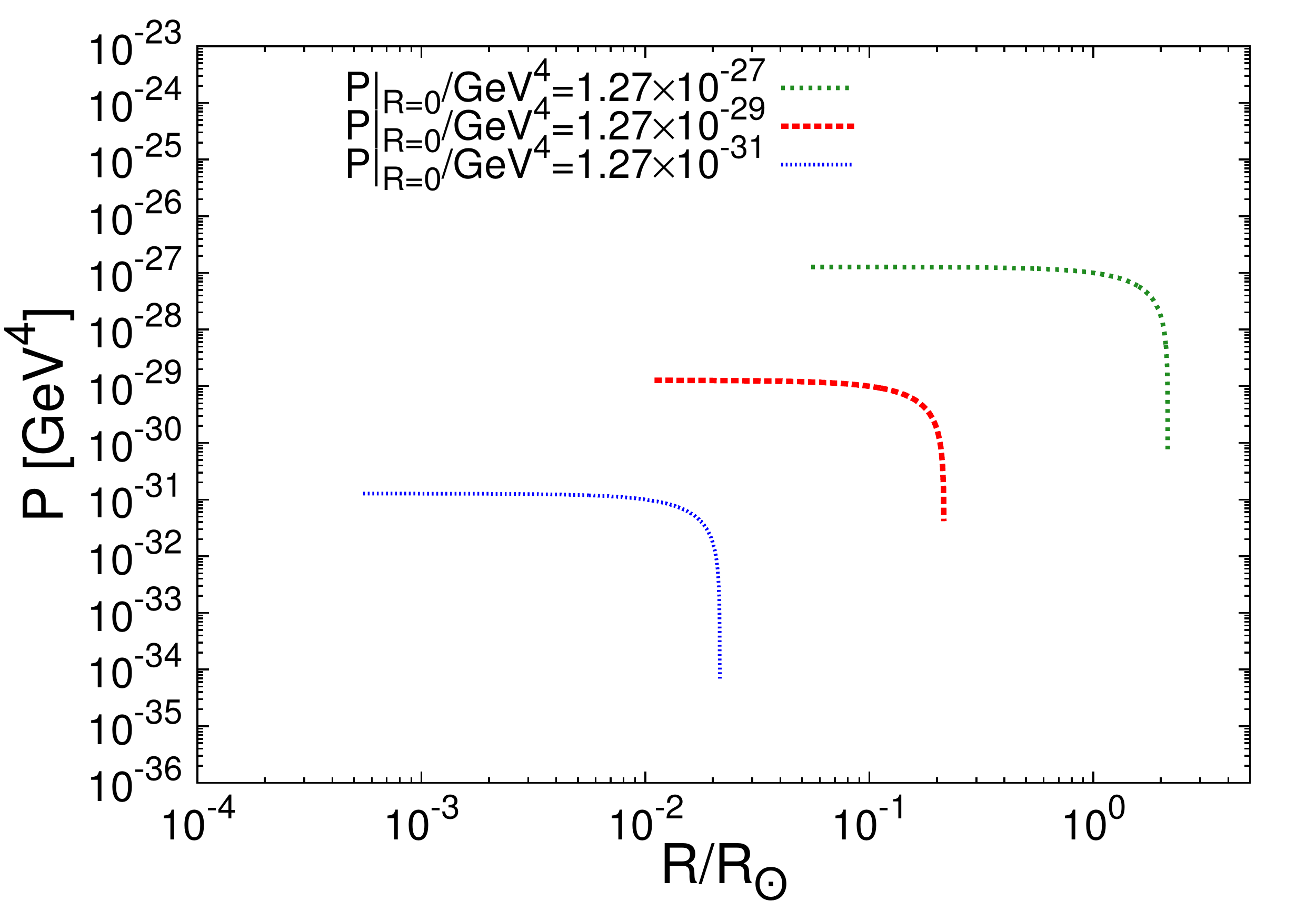}
\includegraphics[height=2.9in,angle=270]{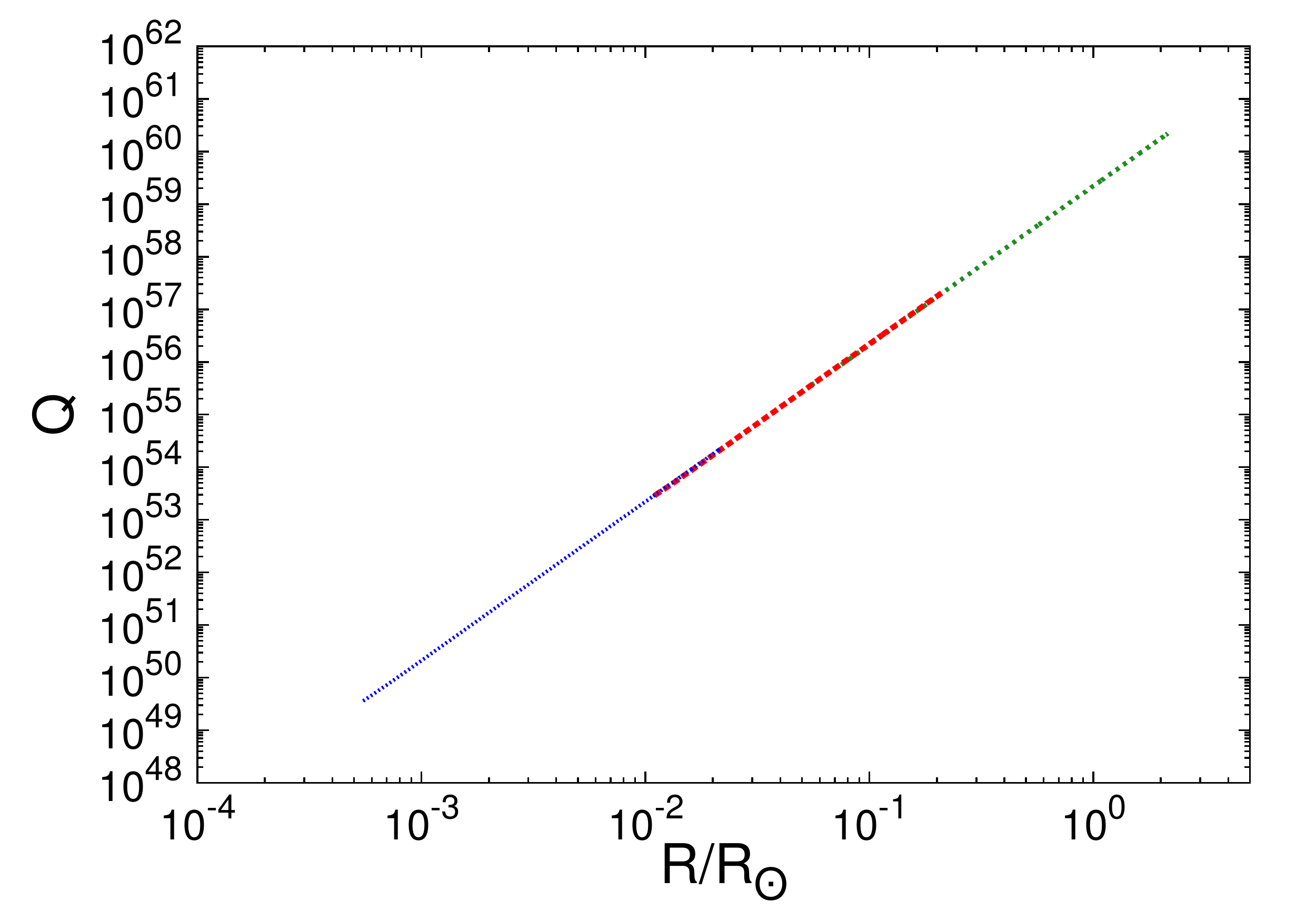}
\includegraphics[height=2.9in,angle=270]{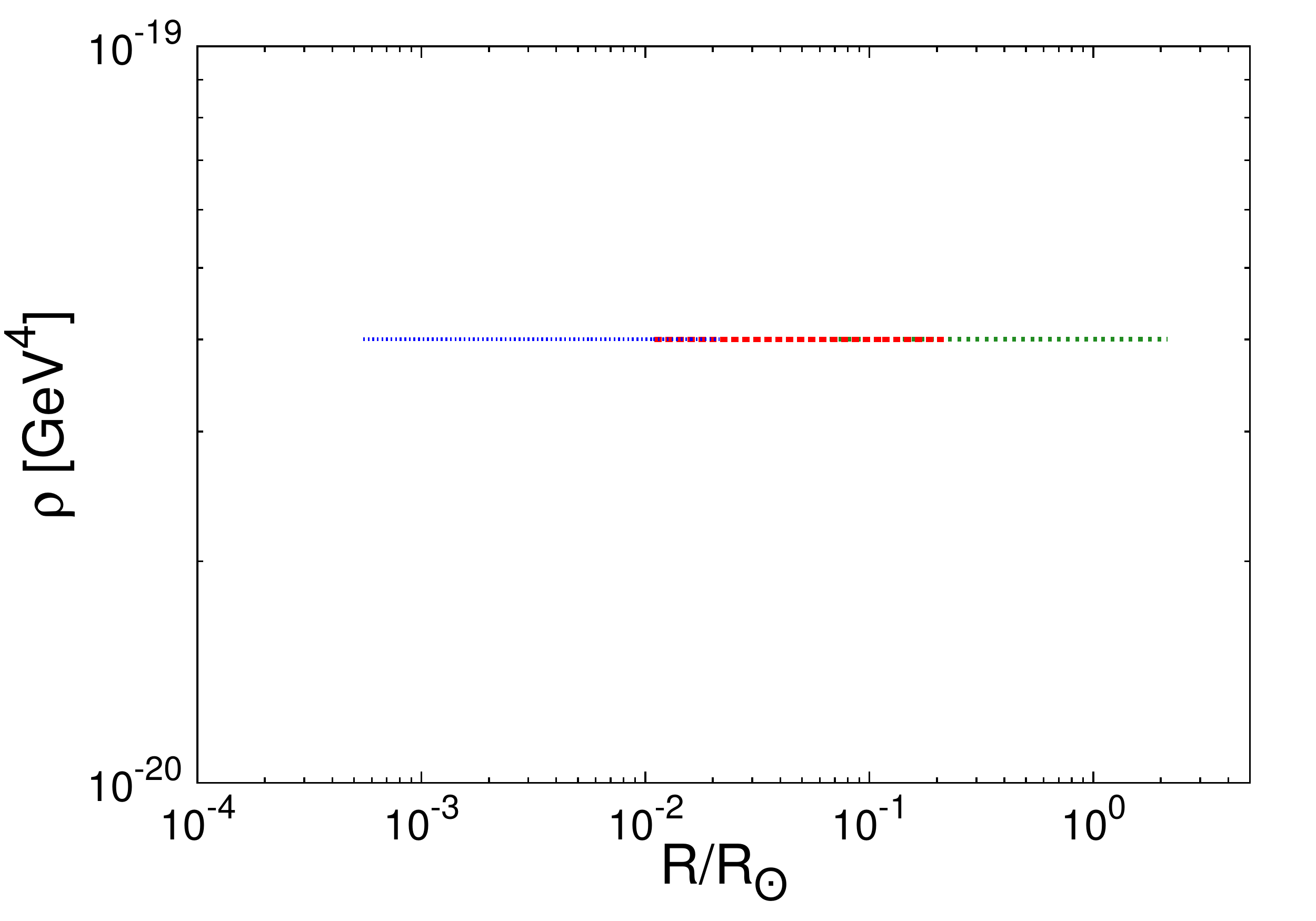}
\caption{\small \label{fig:FB_density}
The pressure $P$ (upper-left), $Q$-charge within radius $R$ (upper-right), and energy density profile (bottom) of a FB with $B^{1/4}=10~{\rm keV}$ for
 three boundary conditions, $P|_{R=0}=1.27\times 10^{-27}~{\rm GeV^4}$,
$P|_{R=0}=1.27\times 10^{-29}~{\rm GeV^4}$,
and $P|_{R=0}=1.27\times 10^{-31}~{\rm GeV^4}$.
Correspondingly, $(M_{\rm FB}/M_{\odot},R_{\rm FB}/R_{\odot})=(6.5079\times 10^{-2},2.149)$, 
$(6.4911\times 10^{-5},0.2149)$, and $(6.5079\times 10^{-8},2.149\times 10^{-2})$.
}
\end{figure}

\begin{figure}[t!]
\centering
\includegraphics[height=3.6in,angle=270]{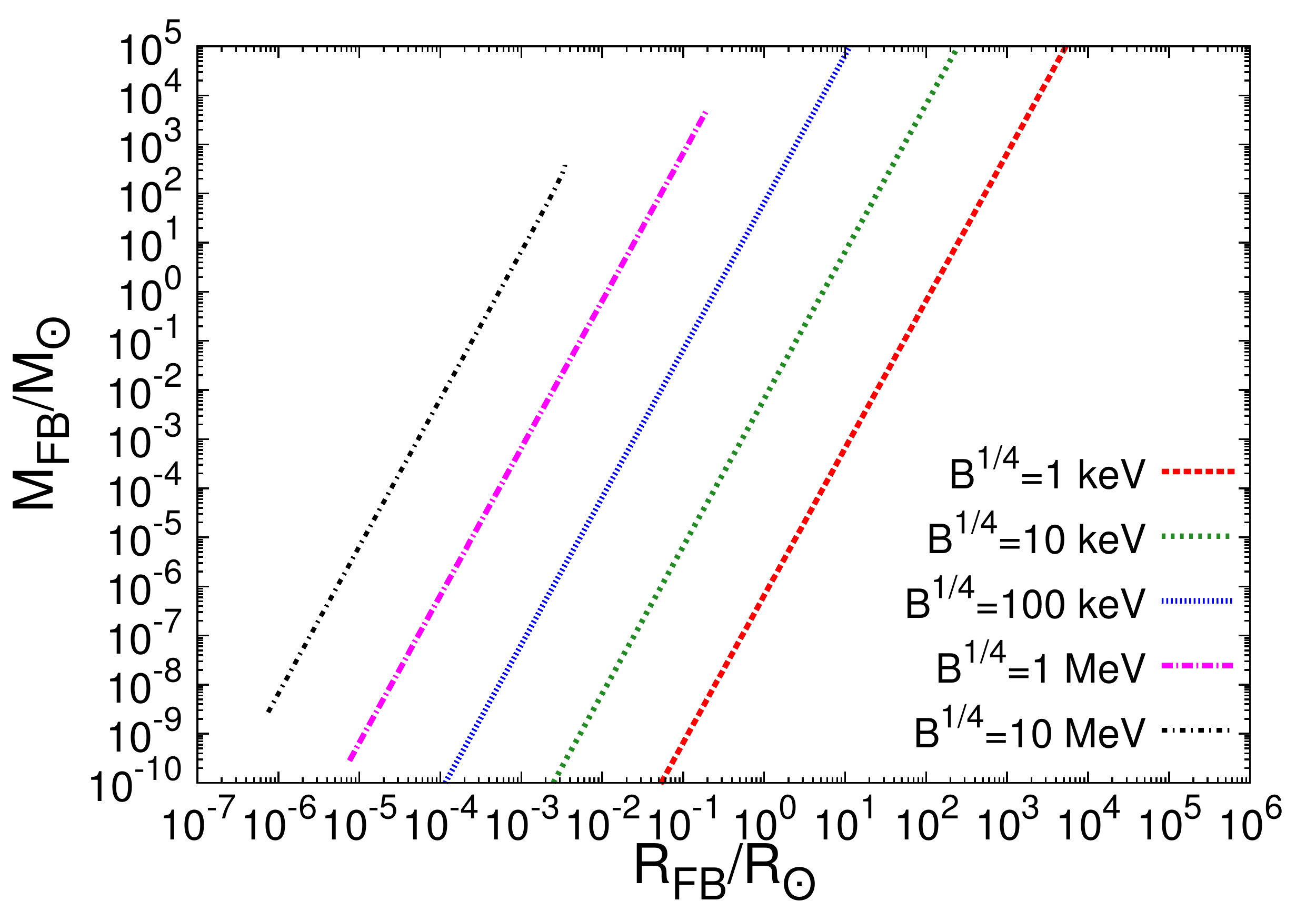}
\caption{\small \label{fig:FB_scan}
Correlation between $M_{\rm FB}$ and $R_{\rm FB}$ obtained by varying $P|_{R=0}$  and solving the  Tolman-Oppenheimer-Volkoff  equation for several values of $B^{1/4}$.
}
\end{figure}

The pressure as a function of $n_{\chi}$ is derived from
\begin{eqnarray}
P = n^2_{\chi} \frac{d\left( \rho_{\rm FB}/n_{\chi} \right)}{dn_{\chi}}
  = \frac{1}{3}\left( \frac{9\pi}{8} \right)^{2/3} n^{4/3}_{\chi}-B\,.
\end{eqnarray}
Using this relation, we obtain the density profile of the FB by solving the TOV equation
with a boundary condition for the pressure at the center of the FB, $P|_{R=0}$.
The result is shown in Fig.~\ref{fig:FB_density}
for three values of $P|_{R=0}$. (In comparison, the pressure at the center of the sun is $1.27 \times 10^{-21}$~GeV$^4$.) From the bottom-right panel,
we see that the FB has a uniform density profile. This is because the constant $B$-dependent term
in Eq.~(\ref{eq:FB_E}) dominates the total energy.
The correlation between $M_{\rm FB}$ and $R_{\rm FB}$ obtained by varying $P|_{R=0}$ is displayed in Fig.~\ref{fig:FB_scan}.

The mass $M_{\rm FB}$ and  radius $R_{\rm FB}$  of a FB are obtained by minimizing the FB energy with respect to
its radius~\cite{Hong:2020est}:
\begin{eqnarray}
\label{eq:FB_mass_radius}
M_{\rm FB}&=& Q_{\rm FB}(12\pi^2 B)^{1/4} \,, \nonumber \\
R_{\rm FB}&=& Q^{1/3}_{\rm FB}
\left[ \frac{3}{16} \left( \frac{3}{2\pi} \right)^{2/3} \frac{1}{B} \right]^{1/4}\,.
\end{eqnarray}
%
The FB relic abundance in the present universe 
is given by~\cite{Hong:2020est}
\begin{eqnarray}
\label{eq:DM_relic}
\Omega_{\rm FB}h^2 &=& \frac{M_{\rm FB}\, n_{\rm FB}|_0 }{3 M^2_{\rm Pl}(H_0/h)^2}\,,
\end{eqnarray}
where the Hubble constant, $H_0=2.13h\times 10^{-42}$~GeV.
We can determine $M_{\rm FB}$ by fixing the value of $\Omega_{\rm FB}h^2$.

 
In Table~\ref{tab:BP}, we list benchmark points that satisfy $\Omega_{\rm FB}h^2 \le 0.12$.
Note that as $B^{1/4}$ gets larger, $M_{\rm FB}$ gets smaller, in apparent contradiction with 
Eq.~(\ref{eq:FB_mass_radius}). The reason for this is as follows. 
Using the saddle point approximation to perform the integral in Eq.~(\ref{eq:F_t}),
the percolation condition, $F(t_\star)=1/e$, is~\cite{Turner:1992tz}
\begin{eqnarray}
\label{eq:approximation}
 8 \pi v^3_w \Gamma(T_\star)\beta^{-4} \simeq 1\,.
\end{eqnarray}
Because $\beta/H_\star$ is roughly constant for 
fixed values of $\lambda$, $A$, and $D$ (see Table~\ref{tab:BP}),
we find $\beta \propto H_\star \propto T^2_\star$. 
Combining with Eq.~(\ref{eq:approximation}), we obtain
\begin{eqnarray}
T^{-4}_\star\, e^{-S_3(T_\star)/T_\star}
\simeq B^{-1}\, e^{-S_3(T_\star)/T_\star} 
\simeq {\rm constant}\,,\ \ \ {\rm{i.e.,}}\ \ \ e^{-S_3(T_\star)/T_\star}  \propto B \,,
\end{eqnarray}
which is not surprising since the bubble nucleation rate per unit volume should grow with the available vacuum energy density.
From Eq.~(\ref{eq:nFB}), $n_{\rm FB}|_0\propto e^{-3/4 \cdot ( S_3(T_\star)/T_\star)}$, so for 
a fixed value of $\Omega_{\rm FB}h^2$, we see that $M_{\rm FB}\propto 1/n_{\rm FB}|_0 \propto e^{3/4 \cdot (S_3(T_\star)/T_\star)} \propto B^{-3/4}$. 


\begin{table}[t]
\caption{\small \label{tab:BP}
Benchmark points with $A=0.1$. 
$N_{\rm events}$ is the number of microlensing events expected in 70 hours of
observation of M31 by Subaru-HSC. 
}
\begin{adjustbox}{width=\textwidth}
\begin{tabular}{c|cc|cc|cc|cc}
\hline
\hline
    & {\bf BP-1} & {\bf BP-2} & {\bf BP-3} & {\bf BP-4} & {\bf BP-5} & {\bf BP-6} & {\bf BP-7} & {\bf BP-8} \\
\hline
\hline
$\lambda$     
            & 0.134 & 0.158 & 0.193 & 0.078 & 0.062 & 0.072 & 0.053 & 0.060  \\
$B^{1/4}/{\rm keV}$     
            & 2.42 & 43.5
            & 34.9 & 64.2
            & 63.6 & 73.2 
            & 284 & 1390 \\
$C/{\rm keV}$    
            & 0.059 & 6.234 & 4.988 & 3.080 & 0.315 & 0.586 & 0.342 & 7.713 \\
$D$     
            & 5.807 & 0.451 & 0.720 & 0.445 & 0.257 & 0.293 & 0.584 & 0.706  \\
$\eta_\chi$     
            & $7.34\times 10^{-6}$ & $1.37\times 10^{-7}$
            & $3.51\times 10^{-6}$ & $4.55\times 10^{-8}$
            & $6.98\times 10^{-9}$ & $3.64\times 10^{-9}$ 
            & $8.54\times 10^{-9}$ & $2.40\times 10^{-8}$ \\
\hline
$T_{{\rm SM}\star}/{\rm keV}$     
            & 1.41 & 100.0 & 64.5 & 128.1 & 164.8 & 169.5 & 427.8 & 1601 \\
$T_\star/{\rm keV}$     
            & 0.57 & 34.2 & 21.6 & 52.3 & 84.8 & 86.9 & 201.0 & 879.0 \\
$T_f/{\rm keV}$     
            & 0.63 & 41.4 & 25.9 & 64.4 & 92.9 & 92.5 & 233.2 & 1005 \\
$S_3(T_\star)/T_\star$     
            & 189 & 188 & 187 & 186 & 187 & 184 & 177 & 171 \\
\hline
$M_{\rm FB}/M_{\odot}$ 
            & $3.37\times 10^{-6}$ & $1.11\times 10^{-6}$
            & $9.66\times 10^{-6}$ & $1.01\times 10^{-7}$
            & $1.08\times 10^{-8}$ & $1.08\times 10^{-9}$ 
            & $9.66\times 10^{-11}$  & $1.09\times 10^{-11}$ \\
$R_{\rm FB}/R_{\odot}$  
            & 0.529 & $7.77\times 10^{-3}$
            & $2.15\times 10^{-2}$ & $2.09\times 10^{-3}$
            & $1.00\times 10^{-3}$ & $3.86\times 10^{-4}$ 
            & $2.83\times 10^{-5}$ & $1.64\times 10^{-6}$ \\
$Q_{\rm FB}$   
            & $4.70\times 10^{56}$ & $8.62\times 10^{54}$
            & $9.38\times 10^{55}$ & $5.34\times 10^{53}$ 
            & $5.74\times 10^{52}$ & $5.00\times 10^{51}$
            & $1.15\times 10^{50}$ & $2.65\times 10^{48}$ \\
\hline
$\alpha$   
            & $1.63\times 10^{-2}$ & $1.56\times 10^{-2}$
            & $1.70\times 10^{-2}$ & $2.83\times 10^{-2}$ 
            & $2.00\times 10^{-2}$ & $1.24\times 10^{-2}$ 
            & $1.79\times 10^{-2}$ & $2.62\times 10^{-2}$ \\
$\beta/H_\star$    
            & $3.43\times 10^{4}$ & $1.57\times 10^{3}$
            & $3.01\times 10^{3}$ & $2.04\times 10^{3}$ 
            & $1.86\times 10^{3}$ & $2.80\times 10^{3}$ 
            & $4.44\times 10^{3}$ & $5.59\times 10^{3}$ \\
$v_\phi/T_\star$  
            & 3.554 & 4.175
            & 3.958 & 4.889
            & 3.987 & 3.501
            & 4.724 & 4.469 \\
$v_w$                   
            & 0.890 & 0.940
            & 0.937 & 0.946
            & 0.886 & 0.854 
            & 0.923 & 0.916 \\
\hline
$\Omega_{\rm FB} h^2$
            & $1.79\times 10^{-2}$ & $5.81\times 10^{-3}$
            & 0.12 & $2.94\times 10^{-3}$ 
            & $4.56\times 10^{-4}$ & $2.70\times 10^{-4}$ 
            & $2.39\times 10^{-3}$ & $3.38\times 10^{-2}$ \\
$N_{\rm events}$
            & 19.5 & 20.4
            & 29.3 & 38.9 
            & 17.5 & 19.3 
            & 46.1 & 29.1 \\
$\Delta N_{\rm eff}$
            & 0.391 & 0.226
            & 0.248 & 0.394 
            & 0.497 & 0.425
            & 0.261 & 0.408 \\
\hline
\hline
\end{tabular}
\end{adjustbox}
\label{2HDM}
\end{table}

\bigskip

\section{Gravitational microlensing}
\label{sec:microlensing}


%

%
If a gravitational lens, i.e., FB, passes along the line of sight of a background source star, 
the star will appear to brighten and subsequently dim, thereby providing the characteristic
signature of microlensing. Microlensing surveys put strong constraints on macroscopic dark matter candidates 
 including FBs.
Subaru-HSC~\cite{Niikura:2017zjd} has surveyed over $8.7\times 10^7$ stars in the M31 galaxy,
which is 770 kpc away from our galactic center~\cite{Niikura:2017zjd}. 
In the seven hours of observation, only one event of transient brightening was detected.
Other surveys like  EROS/MACHO~\cite{Alcock:1998fx} 
and OGLE~\cite{Niikura:2019kqi},
are sensitive to gravitational lenses with masses and radii larger than relevant to us.

\subsection{Microlensing by Fermi balls}
The Einstein radius of a point-like lens is given by
\begin{eqnarray}
\label{eq:RE}
R_E = \sqrt{\frac{4GM_{\rm FB}}{c^2} \frac{D_L D_{LS}}{D_S}}
=\sqrt{\frac{4GM_{\rm FB}D_S}{c^2}x(1-x)}\,,
\end{eqnarray}
and the Einstein angle is $\theta_E = R_E/D_L$.
Here, $G$ is the gravitational constant, $M_{\rm FB}$ is the mass of the lens,
$D_S$ is the distance from the Earth to the source, $D_L$ is the distance from the Earth to the lens,  $D_{LS} = D_S-D_L$, and $x\equiv D_L/D_S$.
Equation.~(\ref{eq:RE}) often  
gives $R_E\simeq \mathcal{O}(R_\odot)$ (e.g., for benchmark point
{\bf BP-1}), 
which is comparable to $R_{\rm FB}$ and the source sizes in M31.
Therefore, we need to account for the finite size of the lenses and sources. 

We briefly outline the procedure of Ref.~\cite{Croon:2020ouk}
which models microlensing signals from spherically symmetric extended sources by spherically symmetric extended lenses.
Assuming $D_S, D_L \gg R_E$,  lensing takes place in the transverse plane containing the lens, so it is convenient to
describe the geometry in this plane. 
With distances in units of $R_E$, the finite source of radius $R_S$ 
has a projected radius $r_S\equiv xR_S/R_E$,
and the distance between the lens center and a point on the limb of the source is
\begin{eqnarray}
\bar{u}(\varphi)=\sqrt{u^2+r^2_S+2ur_S \cos \varphi}\,,
\end{eqnarray}
where $u$ is the distance between the lens and source centers, and 
$\varphi$ is the angular position of the point on the limb measured from the center of the source.
For the uniform density profile of a FB,
\begin{eqnarray}
\rho_{\rm FB}(R)=\frac{M_{\rm FB}}{4 \pi R^3_{\rm FB}/3} \Theta(R-R_{\rm FB})\,,
\end{eqnarray}
the angular position of the image $t_\varphi$ (in units of $\theta_E$) of the point labelled by $\varphi$  
is determined by solving the lensing equation~\cite{Croon:2020ouk,Croon:2020wpr},
\begin{eqnarray}
\label{eq_len}
\bar{u}(\varphi)=\left\lbrace
\begin{array}{lc}
t_\varphi-{1 \over t_\varphi}[1-(1-{t_\varphi^2 \over r_{\rm FB}^2})^{3/2}]\,, &\ \ \  |t_\varphi|<r_{\rm FB}\,,\\
t_\varphi-{1 \over t_\varphi}\,,                             & \ \ \ |t_\varphi|\geq r_{\rm FB} \,,
\end{array}
\right.
\end{eqnarray}
where $r_{\rm FB}\equiv R_{\rm FB}/R_E$.

%

For each $\varphi$ corresponding to a point on the limb, 
Eq.~(\ref{eq_len}) yields
the positions of the (usually one to two)  images at 
$t_{\varphi,i}$.
Neglecting limb darkening, the magnification of each image is the ratio of the area of the image
to the area of the source  in the lens plane~\cite{len_eq,Croon:2020ouk}:
\begin{eqnarray}
\mu_i &=& \eta \frac{1}{\pi r^2_S}\, 
 \int^{\pi}_0 t_{\varphi,i}^2(\psi)  d\psi \,, 
\end{eqnarray}
where 
\begin{eqnarray}
\psi &=& \tan^{-1} \frac{r_S \sin \varphi}{u+r_S\cos \varphi}
\,,~~~\text{$0\leq \psi \leq \pi$ ~\cite{Montero-Camacho:2019jte}}\,, \nonumber 
\end{eqnarray}
is the angular position of the point on the limb measured from the center of the lens, and 
\begin{eqnarray}
\eta &\equiv & {\rm sign}  \left. \frac{dt_{\varphi,i}^2}{d\bar{u}^2} \right\vert_{\varphi=\pi}  \nonumber
\end{eqnarray} 
is the parity of the image $i$.
The total magnification is obtained by summing over the magnification of each image,
\begin{equation}
\mu_{\rm tot}(u)=\sum_{i} \mu_i(u)\,,
\end{equation}
where we have emphasized the dependence on $u$.
Note that the magnifications of images with opposite parities cancel. By convention,
transient brightening is defined as a microlensing event if $\mu_{\rm tot} \ge 1.34$. This threshold corresponds to
magnification by a point-like lens of a point-like source separated by $R_E$, i.e, $u=1$.

\begin{figure}[t!]
\centering
\includegraphics[height=3.6in,angle=270]{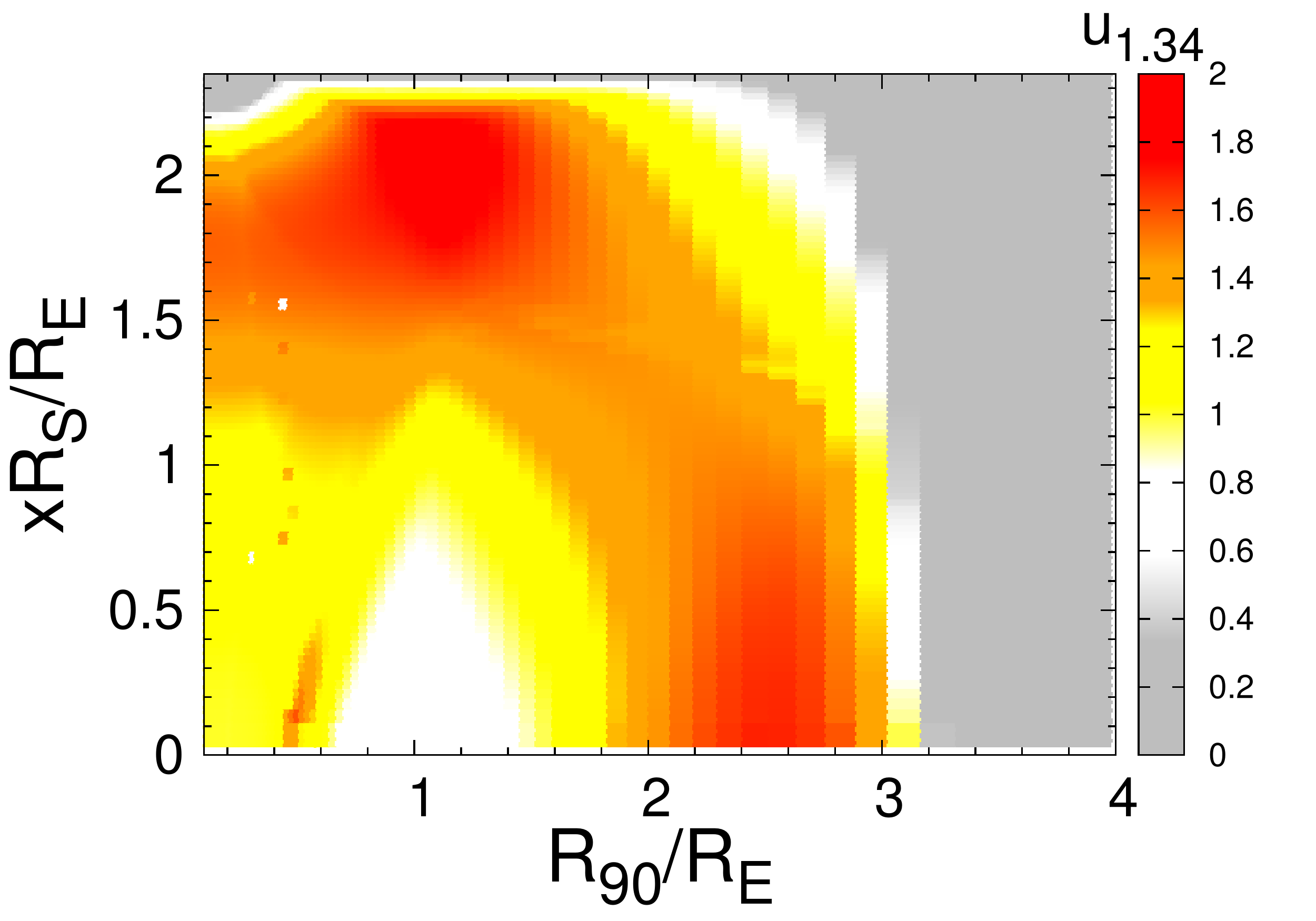}
\caption{\small \label{fig:len_u134_source}
Values of $u_{1.34}$ for lensing by FBs. The $y$-axis is the source radius projected on the lens plane, and 90\% of the FB mass is enclosed within radius $R_{90}$. The distances are in units of $R_E$.
}
\end{figure}

The {\it threshold impact parameter} $u_{1.34}$ is defined as
\begin{eqnarray}
\mu_{\rm tot}(u\leq u_{1.34}) \geq 1.34\,,
\end{eqnarray}
so that the magnification is above threshold for all smaller impact parameters.
Clearly, for a point-like lens
and point-like source, $u_{1.34}=1$.
For the FB density profile,  $u_{1.34}$ is shown in Fig.~\ref{fig:len_u134_source}, where $R_{90} \equiv (0.9)^{1/3} R_{\rm FB}$ is the radius within 
which 90\% of the total mass of the FB is enclosed.



\subsection{Events at Subaru}

If the lenses have a universal mass $M_{\rm FB}$, with velocities from a Maxwell-Boltzmann distribution, 
then the differential event rate per source star is given by~\cite{Griest:1990vu}
\begin{eqnarray}
\frac{d^2\Gamma}{dx dt_E}=D_S {f_{\rm DM} \over M_{\rm FB}} \left[\rho^{\rm DM}_{\rm MW}(r_{\rm MW})\frac{v^4_E(x)}{v^2_{\rm MW}} e^{-v^2_E(x)/ v^2_{\rm MW}}+\rho^{\rm DM}_{\rm M31}(r_{\rm M31})\frac{v^4_E(x)}{v^2_{\rm M31}} e^{-v^2_E(x)/v^2_{\rm M31}}\right],
\end{eqnarray}
where $t_E$ is the amount of time for which the magnification is above threshold, $v_E(x)=2u_{1.34}(x)R_E(x)/t_E$,  $f_{\rm DM}$ is the fraction of DM constituted by FBs, and the most probable speed in M31 (MW) is $v_{\rm M31}=250~{\rm km/s}$ ($v_{\rm MW}=220~{\rm km/s}$). We assume the detection efficiency to be independent of $t_E$
and $R_S$ and set it to 50\%~\cite{Niikura:2017zjd}. 
%
The DM halo profiles of M31 and the Milky Way (MW)
are taken to be NFW,
\begin{eqnarray}
&& \rho^{\rm DM}_{\rm MW, M31}(r) = \frac{\rho'_s}{(r/r'_s)(1+r/r'_s)^2}\,, \ \ \ \ {\rm with}\nonumber \\
&& r_{\rm MW} \equiv  \sqrt{R^2_{\rm sol} -2xR_{\rm sol} D_S \cos\ell \cos b + x^2 D^2_s}\,, \nonumber \\
&& r_{\rm M31} \equiv  D_S(1-x)\,.
\end{eqnarray}
For the MW, $\rho'_s=0.184~{\rm GeV/cm^3}$ and the scale radius $r'_s=21.5~{\rm kpc}$, 
and for M31, $\rho'_s=0.19~{\rm GeV/cm^3}$ and the scale radius $r'_s=25~{\rm kpc}$~\cite{Klypin:2001xu}.
$R_{\rm sol}=8.5~{\rm kpc}$ is the distance from center of the MW to the Sun. The distance between M31 and 
MW is $D_S=770~{\rm kpc}$ and $(\ell,b)=(121.2^\circ,-21.6^\circ)$
are the galactic coordinates of M31.

\begin{figure}[t!]
\centering
\includegraphics[height=2.9in,angle=270]{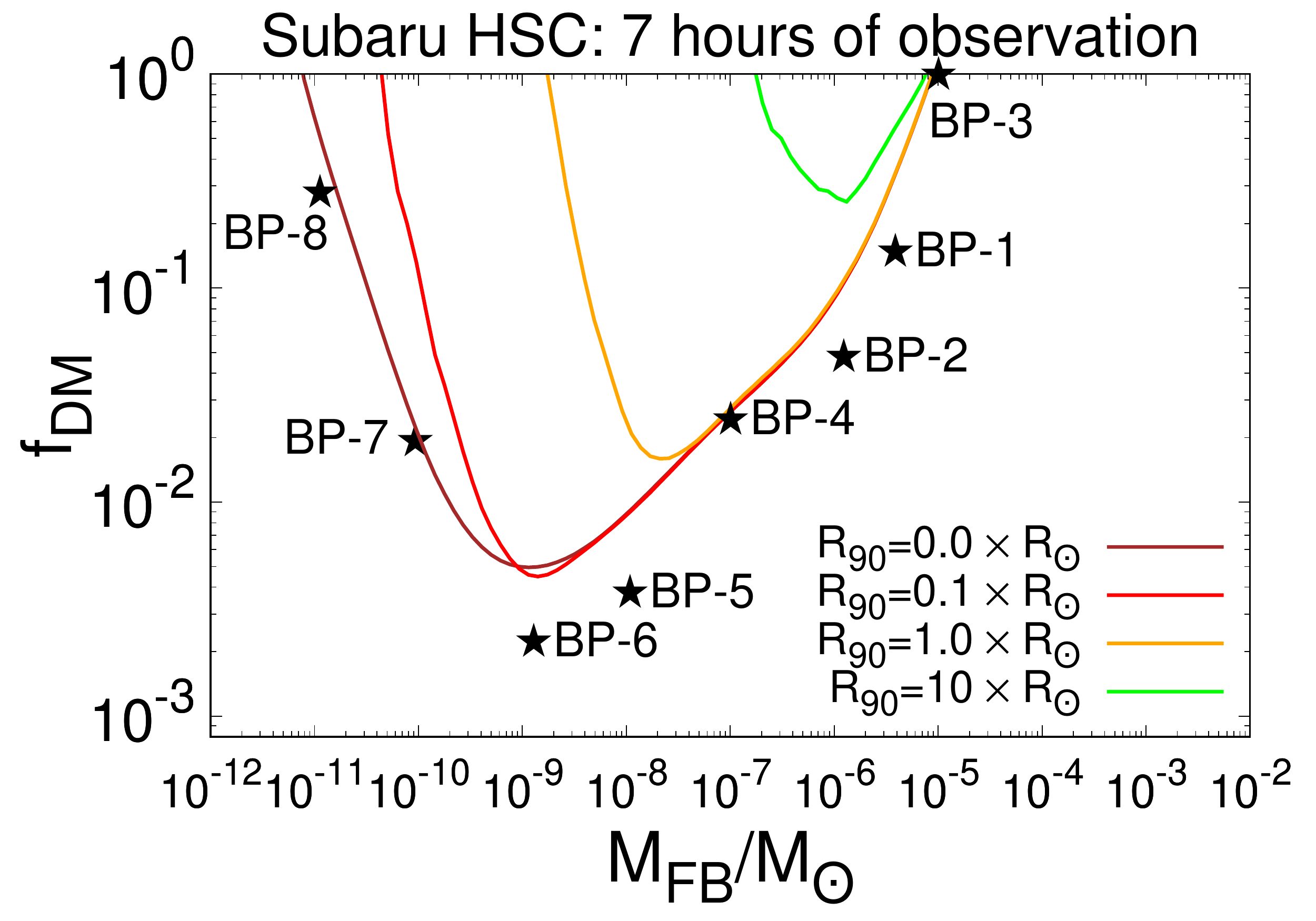}
\includegraphics[height=2.9in,angle=270]{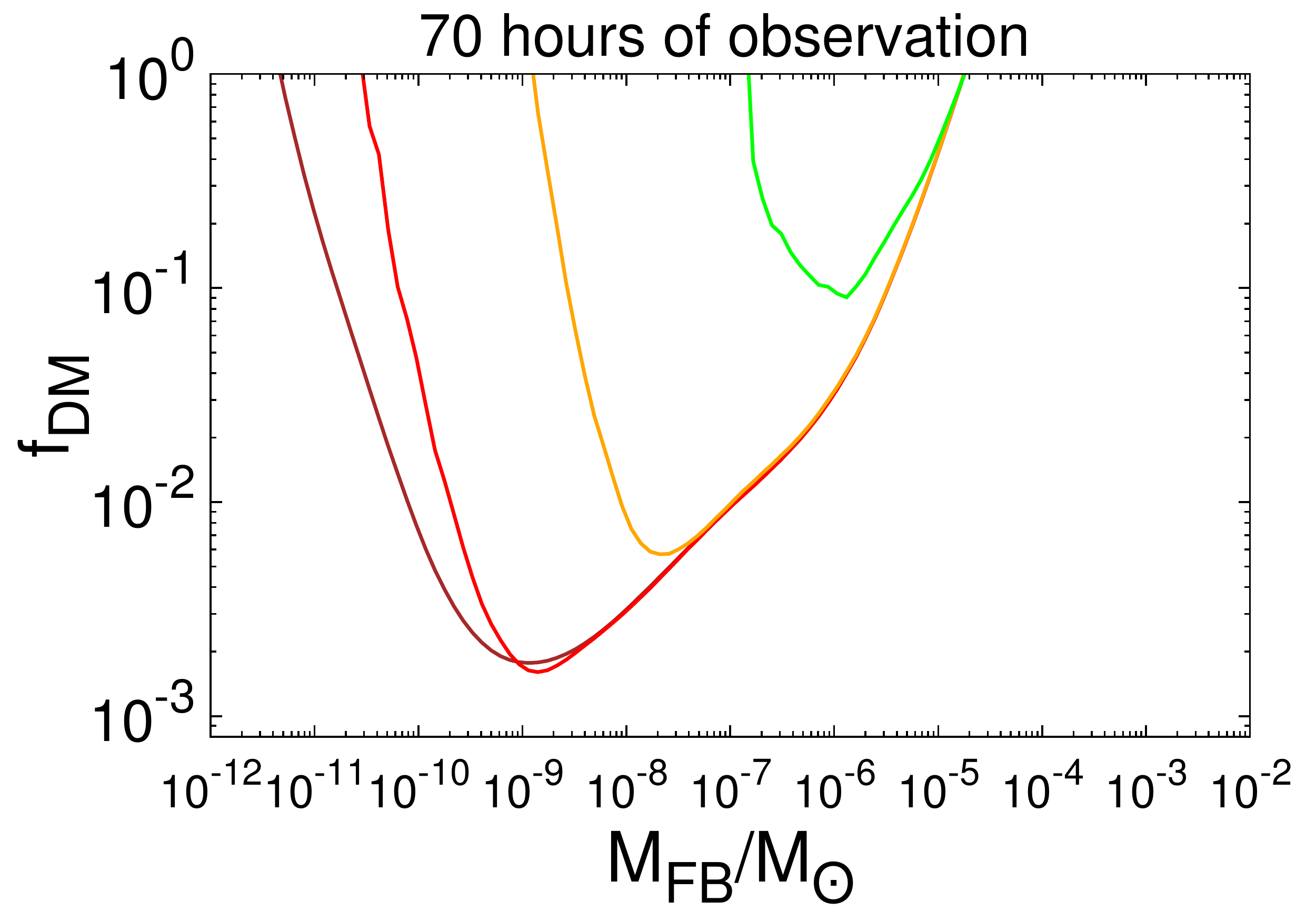}
\caption{\small \label{fig:subaru_fFB}
Left-panel: current 95\% C.L. upper limits on the FB fraction of dark matter 
 from the Subaru-HSC survey of M31. The benchmark points in Table~\ref{tab:BP} are marked with stars.
Right-panel: expected sensitivity of Subaru-HSC with 10 times the observation period.
}
\end{figure}

Taking the stellar radius distribution $dn/dR_S$ in M31 into account~\cite{Smyth:2019whb},
the total number of microlensing events expected  at Subaru-HSC is
\begin{eqnarray}
N_{\rm events}= N_S T_{\rm obs} \int dt_E \int dR_S 
\int^1_0 dx \frac{d^2 \Gamma}{dx dt_E} \frac{dn}{dR_S}\,,
\end{eqnarray}
where $N_S=8.7\times 10^7$ is the number of stars in the survey and $T_{\rm obs}=7~{\rm hrs}$ is the total period of observation.
The left-panel of Fig.~\ref{fig:subaru_fFB} shows the 
95\% C.L. upper limit on $f_{\rm DM}$
by requiring $N_{\rm events} \le 4.74$, corresponding to the one observed event at Subaru.
According to the left-panel of Fig.~\ref{fig:subaru_fFB},
in the point-like lens limit $R_{90}\ll R_\odot$, current Subaru-HSC data constrain 
FBs with $10^{-11}M_\odot \lesssim M_{\rm FB}\lesssim 10^{-5}M_\odot$.

Subaru-HSC plans to monitor M31 for 10 times the current observation period:
$T_{\rm obs}=10\times 7~{\rm hrs}$~\cite{Niikura:2017zjd}. 
Assuming that the observed number of events is proportional to the period of observation, i.e., that 10 events are observed, and the detection efficiency 
is 50\%,
we estimate the 95\% C.L. sensitivity by requiring $N_{\rm events} \le 16.96$; see the right-panel of 
Fig.~\ref{fig:subaru_fFB}.

\section{Gravitational waves}
\label{gws}

\begin{figure}[t!]
\centering
\includegraphics[height=3.6in,angle=270]{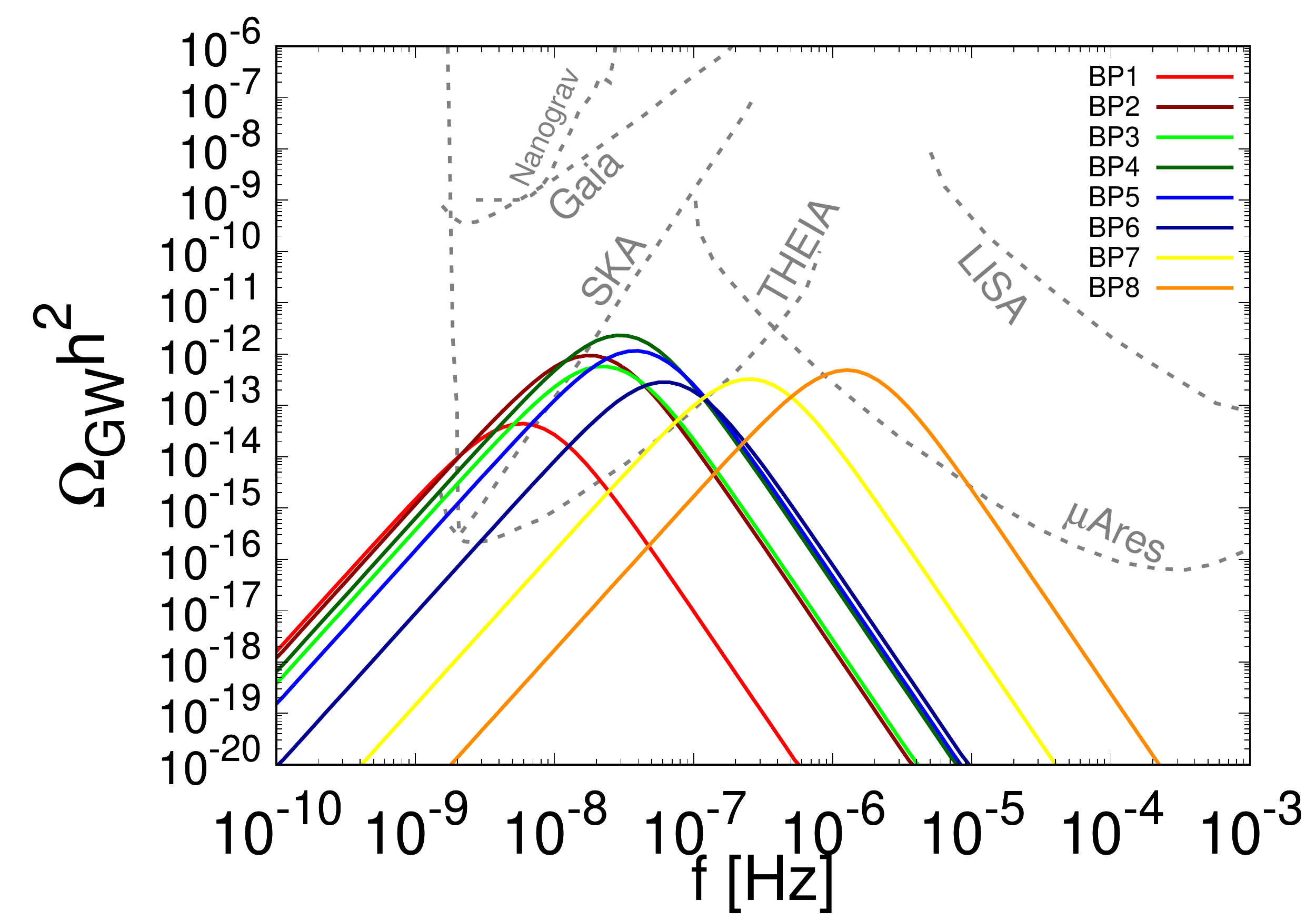}
\caption{\small \label{fig:GW}
The gravitational wave power spectra for the benchmark points in Table~\ref{tab:BP} labeled in order of increasing peak frequency. 
}
\end{figure}

These benchmark points also provide gravitational wave signals because of the FOPT. Other than our assumption that the conversion of latent
heat to GWs is fully efficient, we follow the procedure
used in Ref.~\cite{Marfatia:2020bcs}, which relies on the semi-analytic treatment in Refs.~\cite{Huber:2008hg,Espinosa:2010hh,Caprini:2015zlo}. 
The GW power spectra are shown in Fig.~\ref{fig:GW}.
THEIA is sensitive to all benchmarks points except {\bf BP-8}, which can be detected by $\mu$Ares. 
%

To study the complementarity and correlation between microlensing and GW signals, we perform
a parameter scan in the ranges, $0.05\leq \lambda \leq 0.2$,
$ 1 \leq  B^{1/4}/{\rm keV} \leq 2 \times 10^{3}$,  $0.01 \leq C/{\rm keV} \leq 10$, $0.01 \leq D \leq 10$, and $0.3 \leq T_\star/T_{{\rm SM}\star} \leq 0.6$. We fix $A=0.1$, which
avoids supercooling and gives values of $S_3/T$ for which a FOPT occurs.
For $B^{1/4} < 1~{\rm keV}$, the GW spectra have peak frequencies
 below the sensitivity of SKA/THEIA; see Fig.~\ref{fig:GW}.
For $B^{1/4} > 10^{3}~{\rm keV}$, microlensing cannot be observed because 
$M_{\rm FB}/M_\odot \lesssim \mathcal{O}(10^{-13})$; see Fig.~\ref{fig:subaru_fFB}.
The parameters must satisfy the condition that $T_c$ be real: $\lambda D \geq  A^2- (C/T_0)^2$~\cite{Kehayias:2009tn}. 
Note that the benchmark points in Table~\ref{tab:BP} 
are selected from this scan.

\begin{figure}[t!]
\centering
\includegraphics[height=1.1in]{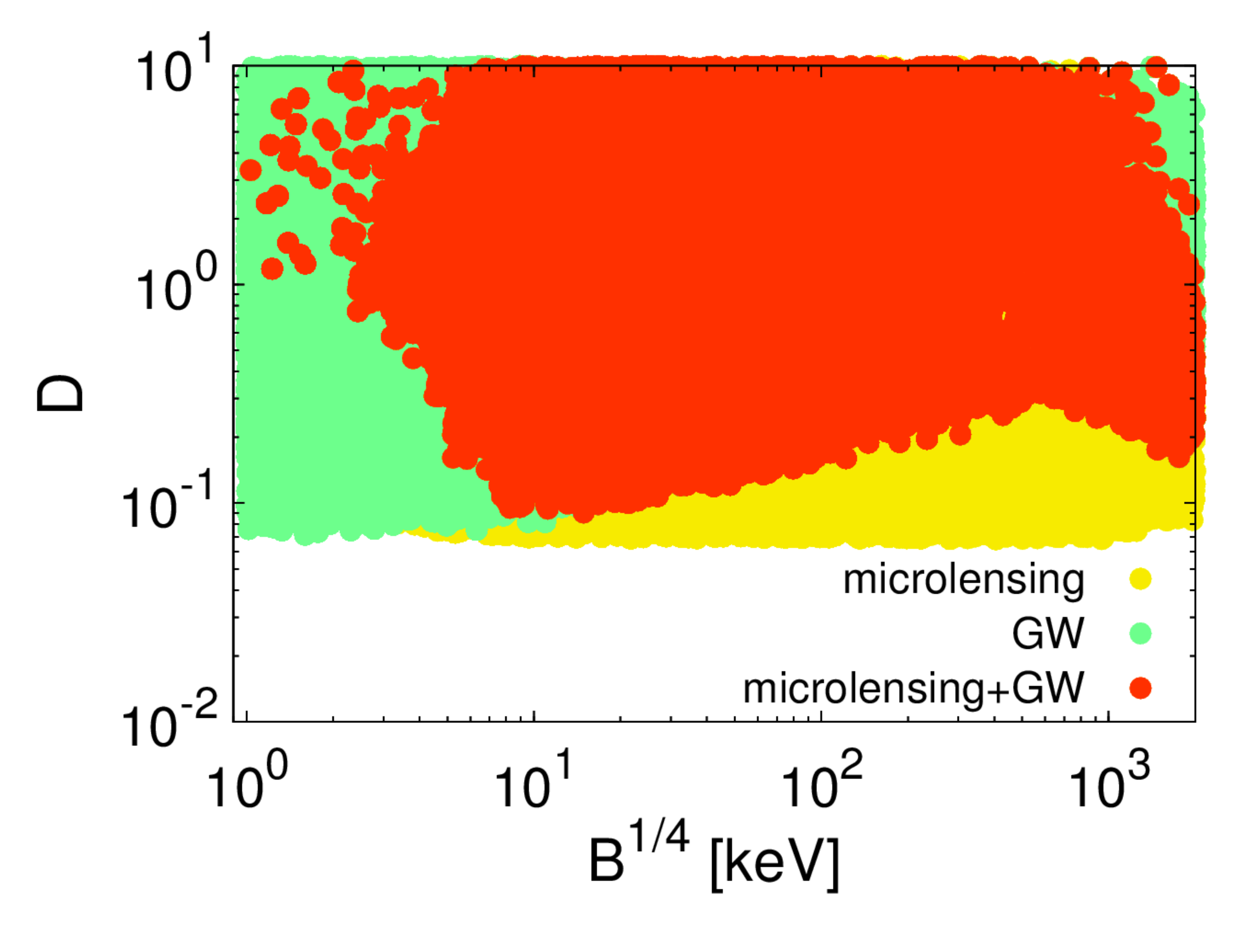}
\includegraphics[height=1.1in]{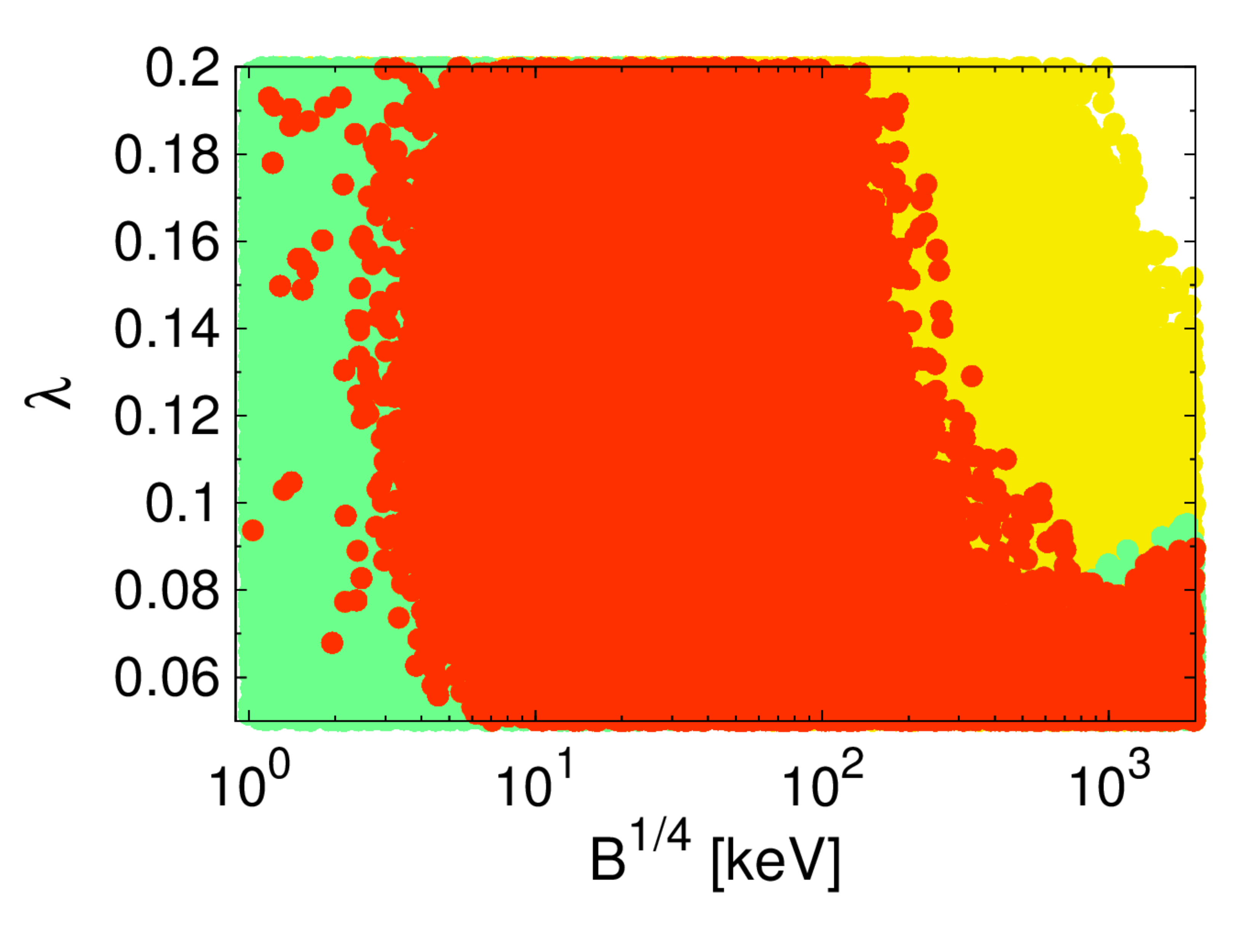}
\includegraphics[height=1.1in]{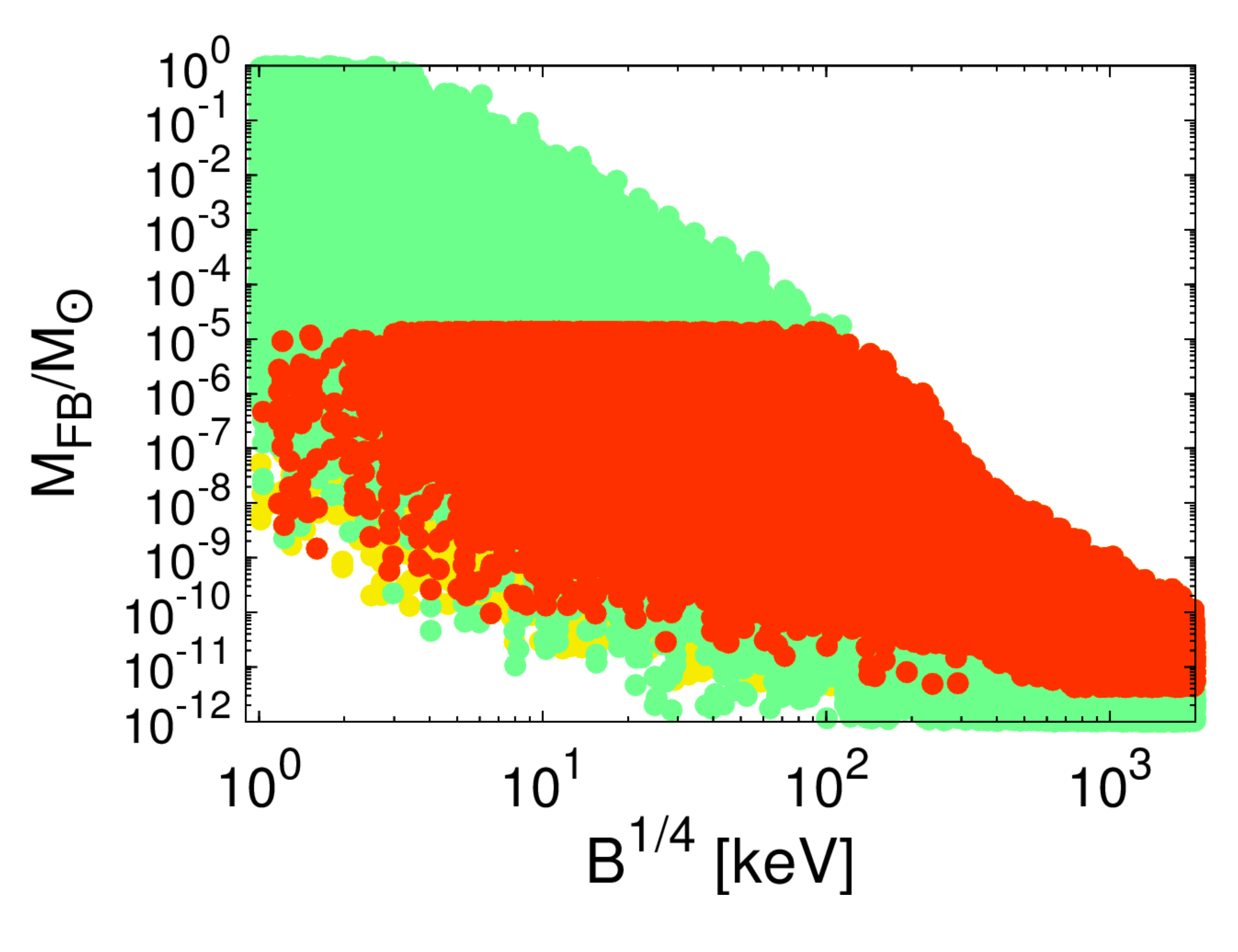}
\includegraphics[height=1.1in]{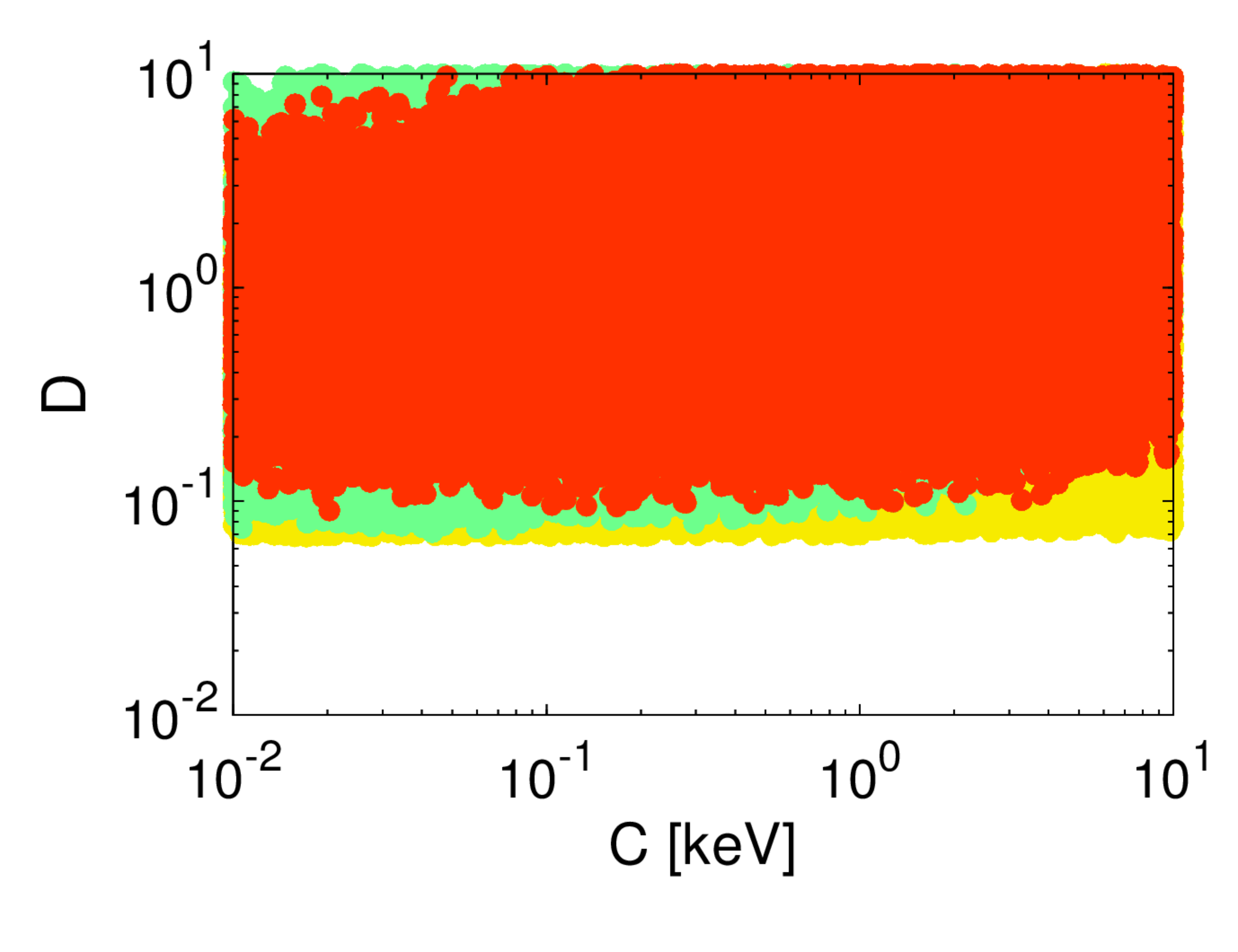}
\includegraphics[height=1.1in]{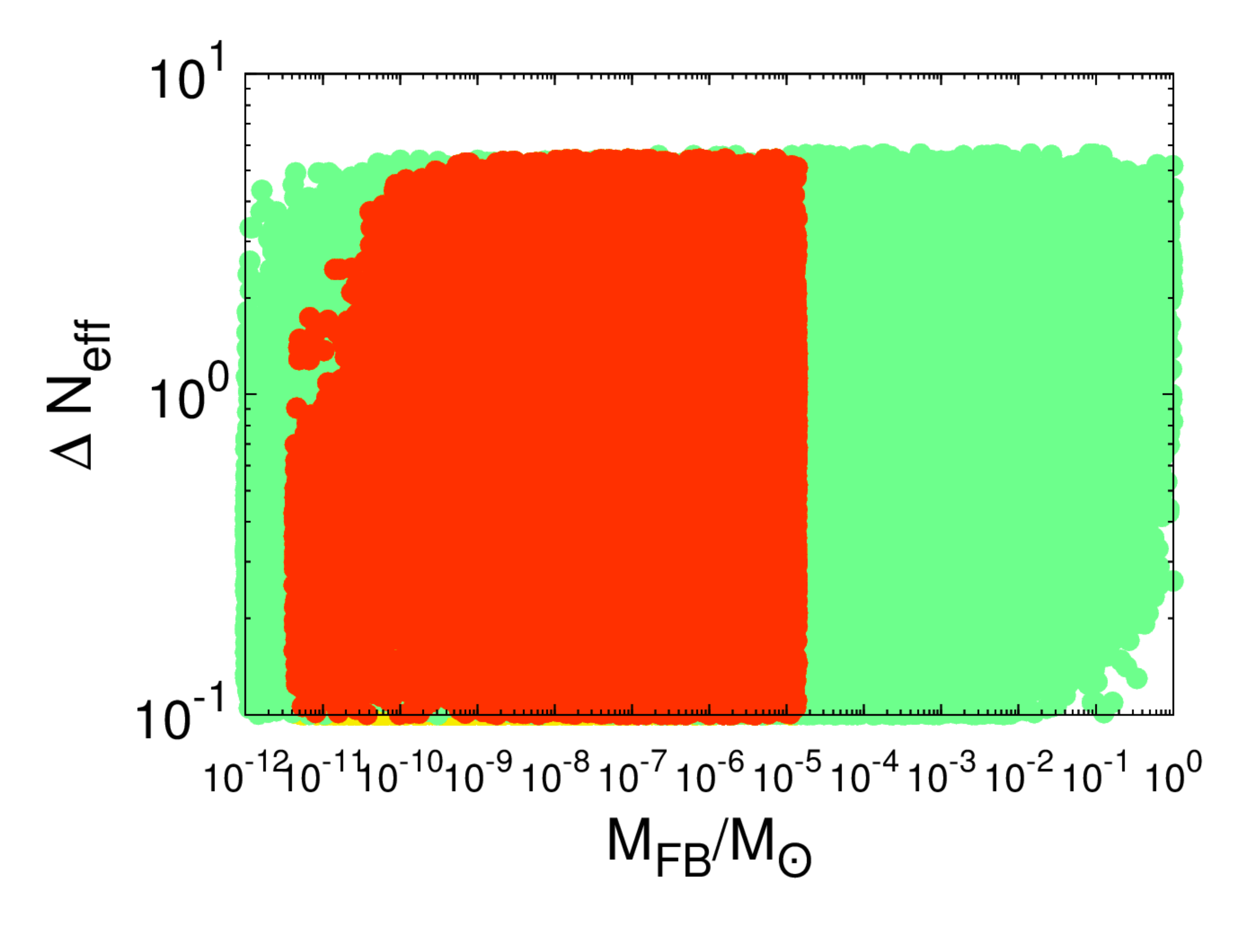}
\includegraphics[height=1.1in]{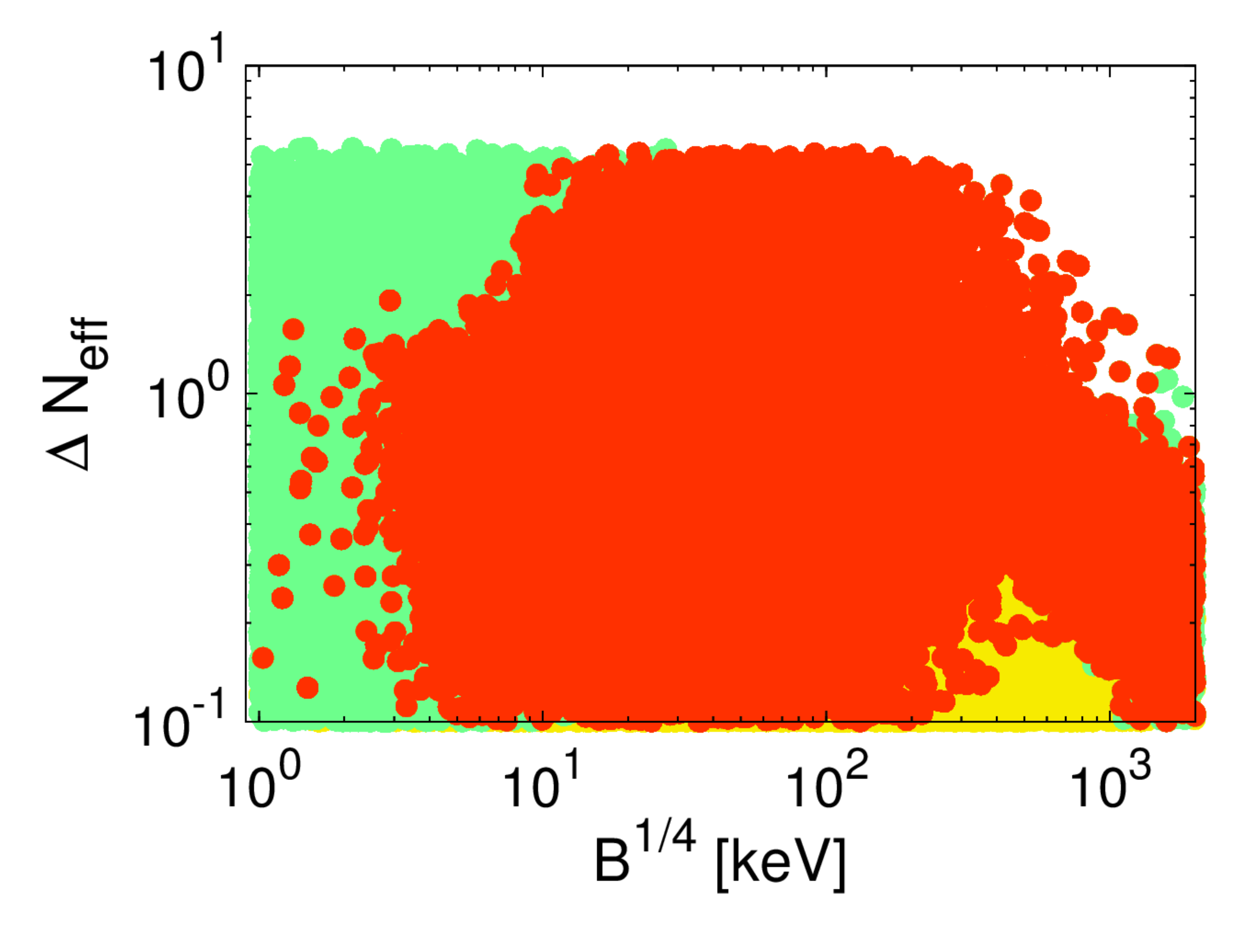}
\includegraphics[height=1.1in]{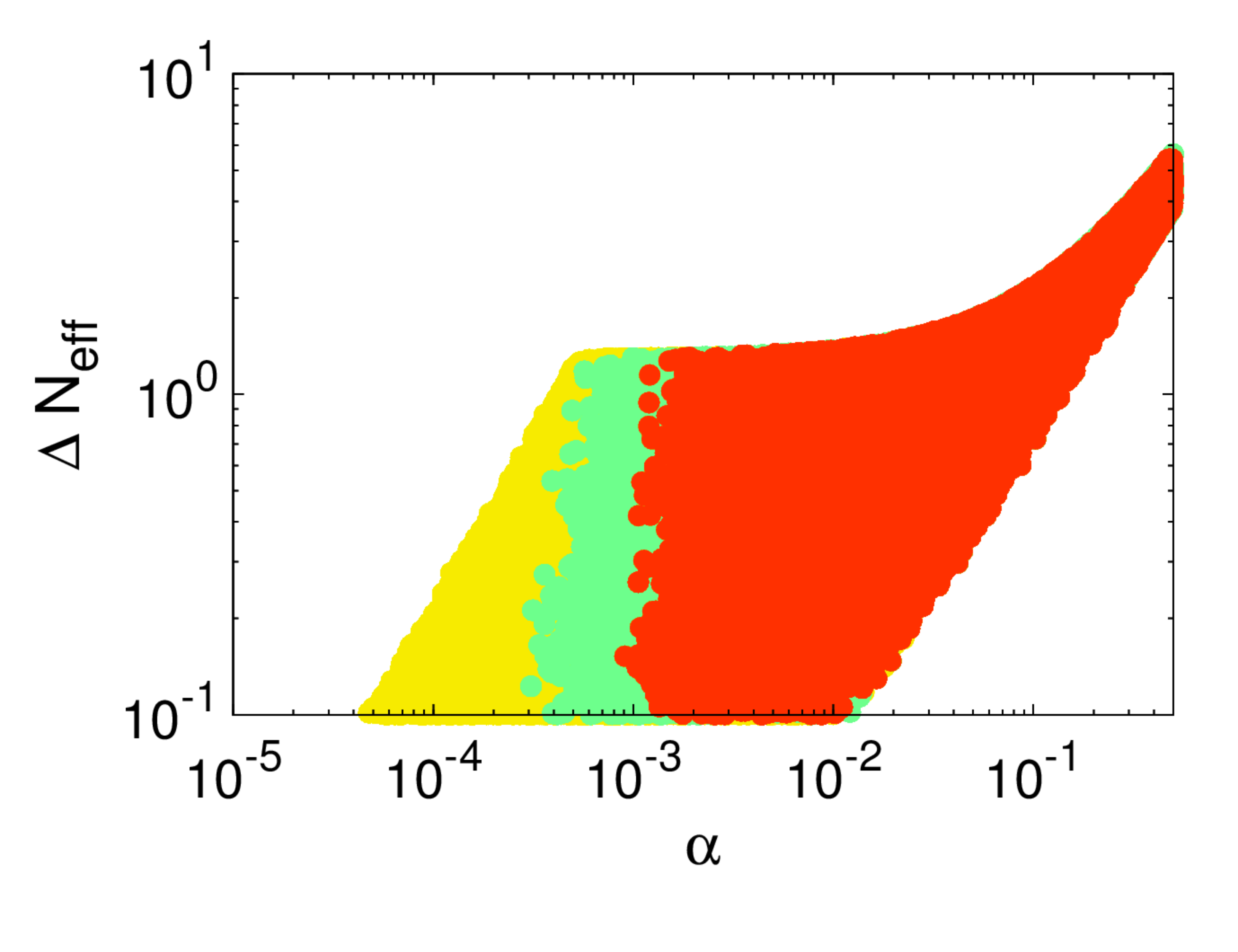}
\includegraphics[height=1.1in]{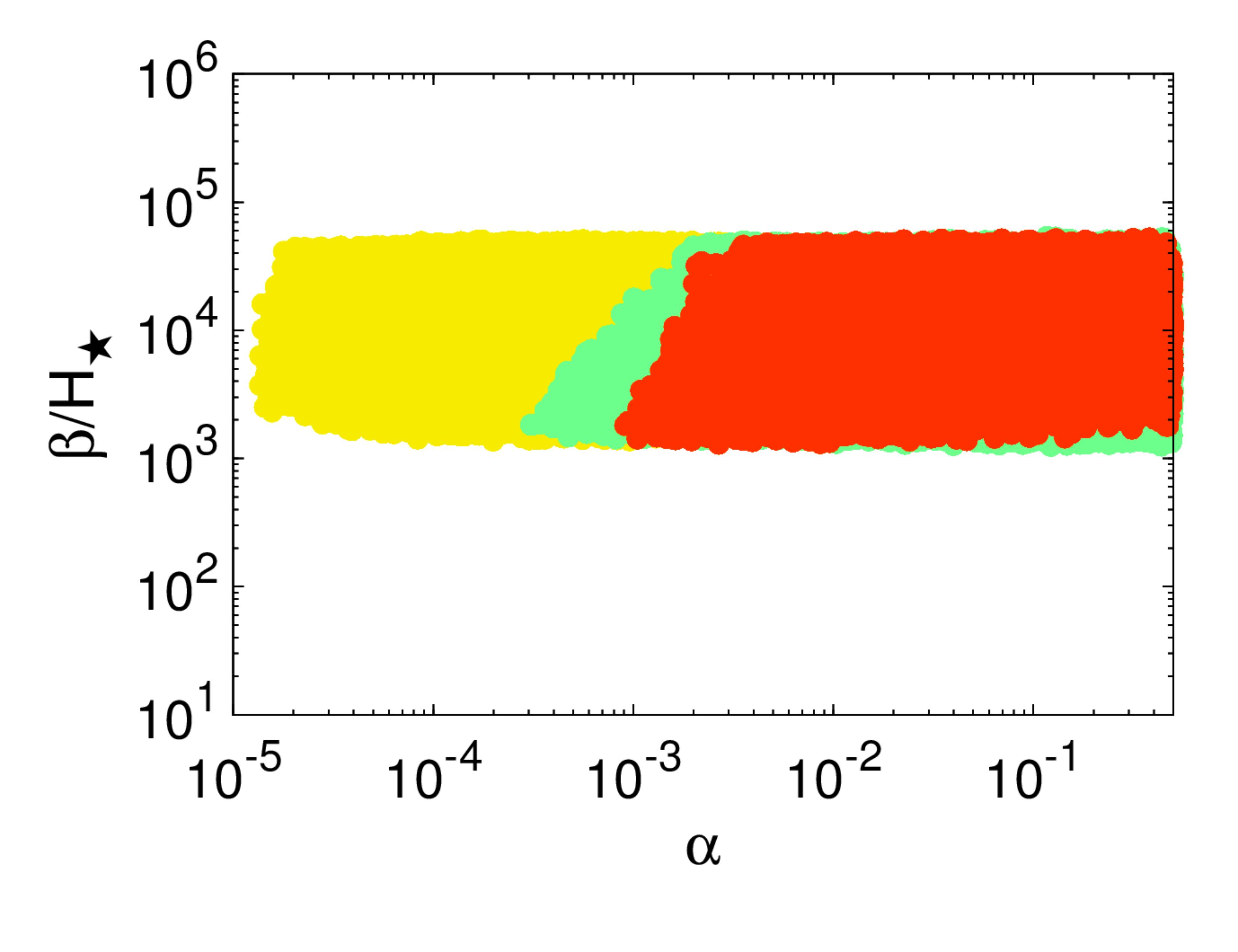}
\caption{\small \label{fig:micro_GW}
The regions of parameter space 
that yield a microlensing signal at Subaru-HSC
and a gravitational wave signal at SKA/THEIA/$\mu$Ares. 
In the red regions correlated GW and microlensing signals can be detected. 
}
\end{figure}

In Fig.~\ref{fig:micro_GW}, the green regions of the parameter space are only probed by GW experiments
because the FBs are too heavy to produce sufficiently many lensing events.
In the yellow regions GW signals go undetected for frequencies with reduced THEIA and $\mu$Ares sensitivity, 
but microlensing observations provide a complementary probe. 
In the red regions, both GW and microlensing signatures of FBs are detectable with existing and planned experiments.
The vertical edges of the red region in the lower left panel correspond to the values of $M_{\rm FB}/M_\odot$ in the right panel of Fig.~\ref{fig:subaru_fFB} at which all sensitivity to microlensing is lost. As confirmed by Fig.~\ref{fig:micro_GW}, to obtain large GW signals with $\alpha \gsim 0.1$, 
$\Delta N_{\rm eff}$ needs to be larger than about unity, which conflicts with constraints from BBN.

\bigskip

\section{Summary}
\label{sec:summary}

We investigated macroscopic dark matter in the form of Fermi balls produced 
in a first-order phase transition in the early universe.
We considered a generic model in which a dark fermion couples to a dark scalar, 
whose thermal effective potential causes a FOPT. 
We find that FBs can be produced  with a wide range of masses, 
$10^{-13}M_\odot \lesssim M_{\rm FB}\lesssim 10^{-3}M_\odot$,
and radii $10^{-6} R_\odot \lesssim R_{\rm FB} \lesssim 10 R_\odot$
for a vacuum energy scale $\mathcal{O}(1) \lesssim B^{1/4}/{\rm keV \lesssim \mathcal{O}(10^{3})~}$.

FBs behave as gravitational lenses and induce microlensing signals.
Current data from the Subaru-HSC survey of M31 constrain the fraction of dark matter composed of FBs, and future data 
from 10 nights of observation will improve the sensitivity considerably; see Fig.~\ref{fig:subaru_fFB}. 

Gravitational waves created during the FOPT that produced FBs are also detectable. 
Under the assumption that the temperature of the dark sector after the FOPT is such that $\Delta N_{\rm eff} \sim 0.1$, a correlation in the GW signal and the microlensing event rate can be found
 using SKA, THEIA, $\mu$Ares and Subaru-HSC data for
 $3\times10^{-12}M_\odot \lsim M_{\rm FB} \lsim 10^{-5}M_\odot$; see Fig.~\ref{fig:micro_GW}.
For FBs heavier than
$10^{-5}M_\odot$, the number of microlensing events above threshold magnification is too low, and gravitational waves provide a complementary signal.

\bigskip
\section*{Acknowledgements}  
D.M. is supported in
part by the U.S. DOE under Grant No. de-sc0010504.
P.T is supported by National Research Foundation of Korea (NRF-2020R1I1A1A01066413).


\end{document}